\documentclass[aps,pre,onecolumn,groupedaddress,superscriptaddress,showpacs,nofootinbib,notitlepage]{revtex4-1}

\usepackage{makecell} 
\usepackage{hyperref}       
\usepackage{url}            
\usepackage{booktabs}       
\usepackage{amsfonts}       
\usepackage{amssymb}
\usepackage{mathtools}
\usepackage{nicefrac}       
\usepackage{microtype}      
\usepackage{amsmath}
\usepackage{xcolor,colortbl}
\usepackage{hhline}
\usepackage{bbm}
\usepackage{changes}
\usepackage{tabularx}
\usepackage{amsthm}
\usepackage{changes}
\usepackage{comment}
\usepackage{color,soul}
\usepackage{xspace}
\usepackage{multirow}
\usepackage{etoolbox}
\usepackage{lineno}
\usepackage{tcolorbox}
\patchcmd{\section}
  {\centering}
  {\raggedright}
  {}
  {}
\patchcmd{\subsection}
  {\centering}
  {\raggedright}
  {}
  {}

\def\ie{\textit{i.e.}}
\def\eg{\textit{e.g.}}

\makeatletter
\newcommand{\lowersim}[2]{%
  \sbox\z@{$#1<$}%
  \raisebox{-\dimexpr\height-\ht\z@}{$\m@th#1#2$}%
}
\makeatother

\makeatletter
\DeclareRobustCommand\onedot{\futurelet\@let@token\@onedot}
\def\@onedot{\ifx\@let@token.\else.\null\fi\xspace}

\begin{document}
\title{Forecasting drug overdose mortality by age in the United States at the national and county levels}

\author{Lucas B\"{o}ttcher}
\email{l.boettcher@fs.de}
\affiliation{Dept.~of Computational Science and Philosophy, Frankfurt School of Finance and Management, Frankfurt am Main, 60322, Germany}
\author{Tom Chou}
\email{tomchou@ucla.edu}
\affiliation{Dept.~of Computational Medicine, UCLA, Los Angeles, CA 90095-1766}
\author{Maria R. D'Orsogna}
\email{dorsogna@csun.edu}
\affiliation{Dept.~of Mathematics, California State University at Northridge, Los Angeles, CA 91330-8313}
\affiliation{Dept.~of Computational Medicine, UCLA, Los Angeles, CA 90095-1766}
\date{\today}
\begin{abstract}
The drug overdose crisis in the United States continues to
intensify. Fatalities have increased five-fold since 1999
reaching a record high of 108,000 deaths in 2021. The epidemic has
unfolded through distinct waves of different drug types, uniquely
impacting various age, gender, race and ethnic groups in specific
geographical areas.
One major challenge in designing effective interventions is the
forecasting of age-specific overdose patterns at the local level so that
prevention and preparedness can be effectively delivered.
We develop a forecasting method that assimilates observational
data obtained from the CDC WONDER database with an age-structured model of
addiction and overdose mortality.
We apply our method nationwide and to three select areas: Los Angeles
County, Cook County and the five boroughs of New York City, providing forecasts of drug-overdose mortality and estimates of relevant epidemiological quantities, such as mortality and age-specific addiction rates.
\end{abstract}
\maketitle
\begin{tcolorbox}
\textbf{Significance:} The drug overdose epidemic in the United States continues to escalate, with fatalities increasing five-fold since 1999 and reaching a record high of 108,000 individuals in 2021. The crisis is characterized by distinct waves of drug types, disproportionately affecting various demographic groups in specific geographical regions. One major challenge in designing effective interventions is to forecast age-specific overdose patterns to facilitate targeted prevention and preparedness efforts. To this end, we propose a forecasting approach that integrates observational data with an age-structured model of addiction and overdose mortality. Applying this method nationwide and in areas that are highly impacted by the overdose crisis, we provide robust drug-overdose mortality forecasts offering vital insights for effective interventions.
\end{tcolorbox}

The United States is currently experiencing one of its worst
drug crises, with alarming increases in fatal
overdose rates. According to data from the Center for Disease Control
(CDC), over 108,000 persons died from drug overdose in 2021, the
highest number ever recorded in a single year and a 17$\%$ increase
over the previous record high in 2020~\cite{mattson2021trends}.  Most
recent overdose deaths involve synthetic opioids such as fentanyl,
psychostimulants such as methamphetamines and, to a lesser degree,
prescription opioids such as oxycodone, and heroin~\cite{ODonnell2020}.
Many factors may have contributed to this surge, including
  the overall increased supply of synthetic, low-cost
  drugs~\cite{Jones2018, Armenian2018}, the ease with which illegal substances may be
  purchased online~\cite{Forman2006, Mackey2017, Lamy2020}, the uncontrolled mixing of drugs of
  different potency~\cite{Duhart2022, Kariisa2023}, and societal changes leading to
  ``deaths of despair'' \cite{Stein2017, Case2020}. Although these elements have
fueled drug abuse for quite some time, most of them have been
exacerbated by the COVID-19
pandemic~\cite{Friedman2021,dorsogna2023fentanyl}.

Both the CDC and the National Center for Health Statistics have been
systematically collecting information on overdose mortality since
1999, and, according to slightly different classifications, since
1979. The relevant data is publicly accessible through the CDC
Wide-ranging Online Data for Epidemiologic Research (WONDER) portal
which is updated at the end of each calendar year with final data
associated to the prior year, resulting in a one-year lag.  Many
groups have dissected these data by stratifying overdoses according to
substance type, year, age, gender, race, and geography. These studies
have revealed several spatio-temporal ``overdose waves' across the
United States, the emergence of new trends, demographic and
geographical shifts, and social
disparities~\cite{Jalal2018,Peters2020,powell2023trends,dorsogna2023fentanyl}.

While providing up-to-date snapshots and following the course of past
overdose deaths helps shed light on the evolution of the drug
epidemic~\cite{Segel2021}, forecasting future overdose patterns, even
in the short term, would allow for targeted preventive interventions
and ensure the preparedness of public health agencies \cite{Jalal2018, Blanco2022, Lim2022}. Due to
demographic, political and legislative heterogeneities across the
United States, predictions on the national scale would be much less
effective than those made at the more local
level~\cite{Borquez2022}. Analysis at a more ``granular'' scale
retains specific drug-market, socioeconomic, cultural, and
geo-historical conditions that distinctly affect the drug overdose
trajectories.  By not lumping these factors together, more realistic
forecasts and tailored interventions \cite{Monnat2018, Rigg2018} can
be developed. For instance, while drug overdoses may be decreasing at
the regional level, certain counties, urban centers, or even zip codes
within the same region may be experiencing surges among given
sub-populations due to the introduction of new drugs to a
circumscribed market.

As data collecting and manipulation capabilities have expanded,
predicting overdose mortality (at any scale), while still in its
infancy, has become a rapidly growing field.  Given the many aspects
of the drug addiction crisis, current studies rely on a variety of
information including data on past overdoses, hospitalization, arrest,
internet searches, painkiller prescription and drug-seizure rates.
Quantitative tools used in these endeavors include statistical
regression, geospatial analyses, mathematical modeling and machine
learning~\cite{Brownstein2010, Basak2019, Campo2020, Marks2021,
  Marks2021b, Sumetsky2021, Wagner2021, Stringfellow2022}.

In this paper we advance the state-of-the-art in drug overdose
forecasting by combining a mechanistic model describing age-stratified
drug-overdose fatalities with recorded mortalities using data
assimilation techniques \cite{crassidis2004optimal, law2015data,
  bottcher2023forecasting}. The latter were first developed within the
geological and atmospheric sciences to merge high-dimensional
dynamical systems with large datasets to produce weather and climate
forecasts. After decades of continuous improvement to both algorithms
and computing infrastructure, modern operational weather forecasting
centers are able to process about $10^7$ observations per
day~\cite{bauer2015quiet}. In addition to its application in climate dynamics,
data assimilation has been used to estimate parameters in systems
biology~\cite{lillacci2010parameter}, to provide risk-dependent
individual contact interventions during
outbreaks~\cite{schneider2021epidemic}, to identify patients with
antibiotic-resistant bacteria in hospital
wards~\cite{pei2021identifying}, and to quantify the proportion of
undocumented COVID-19 cases~\cite{li2020substantial}. One reason for
the successful integration of mechanistic models with data
assimilation methods across different fields is that the algorithms
are computationally efficient and provide good forecasts even when
training data are sparse~\cite{bomfim2020predicting}. 
Furthermore, since they are coupled to mechanistic models, data assimilation methods allow
to estimate parameters that carry a physical or biological meaning
and to follow their evolution over time. This interpretability, both of the parameters and
of their dynamics, is very valuable for decision-making and
formulating intervention policies.  Finally, contrary to other
techniques, data assimilation methods produce interval estimates and
not just point estimates. This allows to quantify confidence
intervals and accurately assess uncertainties and risks.

The mechanistic model that we use in this work is based the
Kermack--McKendrick theory~\cite{m1925applications,kermack1991contributionsI,kermack1991contributionsII,kermack1991contributionsIII,diekmann2021discrete}
and describes an age-structured population that suffers from 
substance use disorder (SUD). Our model includes population aging, addiction
(\ie, the age-dependent onset of SUD in a certain subset of the
overall population), and drug-induced mortality.  Using data
assimilation to combine our drug-overdose model with data from CDC
WONDER, we develop a forecasting tool for age-stratified drug-overdose
mortality in the United States. In the next section, we illustrate the
basic principles of our method by generating nationwide drug-overdose
mortality forecasts and by extracting the time evolution of
epidemiological quantities such as mortality and addiction rates.  We
compare our predictions with overdose data for select past years and offer
short-term projections for drug overdose mortality.  Similarly, we
generate drug-overdose forecasts for select counties or metropolitan
areas that display a large number of overdose fatalities: Los Angeles
County, CA, Cook County, IL and the five boroughs of New York City. 
Our forecasts show that age-structured
population models combined with data assimilation methods can produce
reliable predictions of drug-overdose deaths both at the national and
county levels.  Our approach and its results can inform early warning
systems, help tailor interventions, and prioritize resources
distribution to areas most impacted by the current drug epidemic.

\section*{Results}
\subsection*{Forecasting overdose fatalities in the United States}
\begin{figure}
    \centering
    \includegraphics{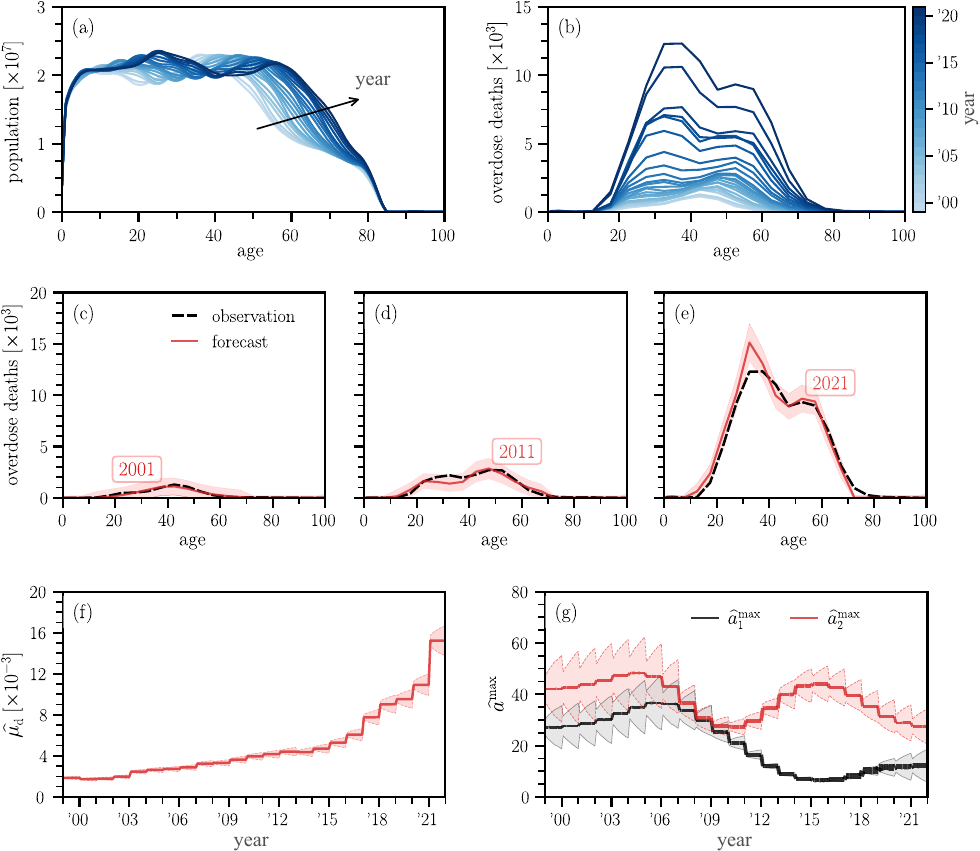}
    \caption{\textbf{Forecasting nationwide overdose fatalities.} (a,b) United
      States population and overdose deaths as a function of age
      (0--100 years) and time (1999--2021). (c--e) Forecasts of
        overdose deaths as a function of age. Solid red curves and red shaded areas
      indicate mean predictions and 3$\sigma$ intervals,
      respectively. Observed fatalities are indicated by dashed black
      curves. Ages are binned in 5 year intervals and prediction values are displayed at the center of each
        bin interval.
      (f) Evolution of estimated drug-caused mortality rate
      $\hat {\mu}_{\rm d}$ and corresponding 3$\sigma$ intervals. (g)
      Evolution of the estimated ages for which the onset of SUD is
      largest and corresponding 3$\sigma$ intervals. To account for
      potential population shifts, we utilize two age-stratified
      influxes peaked at $\hat{a}^{\rm max}_1, \hat{a}^{\rm max}_2$.
      The thickness of the curves is proportional to their respective
      magnitudes $\hat{r}_1$ and $\hat{r}_2$. As can be visually
      inferred, the influx peaked at the younger age $\hat{a}^{\rm
        max}_1$ begins to carry more weight than its counterpart
      peaked at $\hat{a}^{\rm max}_2$ around 2015, indicating a shift
      towards a preponderance of young persons with SUD.  Filter
      updates occur in the beginning of each year.}
    \label{fig:US_forecast}
\end{figure}
The Kermack--McKendrick
model~\cite{m1925applications,kermack1991contributionsI,kermack1991contributionsII,kermack1991contributionsIII}
is a standard tool in mathematical epidemiology and is used to
describe the evolution of an age-structured population with
age-dependent infection and recovery rates.  Related structured
population models have found utility in describing cell populations
\cite{Xia_SIAM}, demographics and birth control policies
\cite{ChildPolicy}, the progression of infectious
diseases~\cite{diekmann2021discrete} such as measles~\cite{SCH1984},
tuberculosis~\cite{CAS1984}, HIV~\cite{RON2007}, and
COVID-19~\cite{bottcher2020case} and more recently 
in the social sciences \cite{CHU2018} and in studies of
drug addiction~\cite{YAN2016,LIU2019,CHE2020,DIN2020,DUA2021,KHA2021}.
In this work, we combine an age-stratified model of overdose
fatalities with corresponding observational data from the CDC WONDER
database, using an ensemble Kalman filter
(EnKF)~\cite{evensen1994sequential} as data assimilation method (see
Online Methods for further information on the age-structured model,
ensemble Kalman filter, and overdose data).

We model both the evolution of individuals with SUD and the number of
fatal drug overdoses across different age classes in yearly
increments. To estimate age-specific addiction rates (\ie, the influx
of new SUD cases in a certain age class per unit time) and mortality
rates, we use the age-structured population data and fatal overdose
data tallied by the CDC WONDER database as inputs to our EnKF.
Figures~\ref{fig:US_forecast}(a,b) show the evolution of the United
States population and overdose fatalities from 1999 (light blue) to
2021 (dark blue).  Within this timeframe, the population of the
country between ages 0--85 rose from 275 to 326 million individuals;
Figure~\ref{fig:US_forecast}(a) shows that the largest increases
occurred between ages 20-40 years and that the age-structured
population distribution is marked by two characteristic peaks: one
arising between ages 20--30 years and the other between ages 40--60
years.  The age-structured fatal overdose distribution in
Fig.~\ref{fig:US_forecast}(b) reveals that between 1999 and 2014, the
largest proportion of overdose deaths occurred between ages 40-50
years. During a second phase, spanning from 2015 to 2019, overdoses
peaked within the 35-40 year age group. A sudden surge in overdose
fatalities is observed beginning in 2020; the onset of this third
phase is concurrent with the advent of COVID-19. These three phases do
not define rigid classifications; rather, they provide reference
points to facilitate data interpretation and guide our analysis.

We use these qualitative observations to guide the development of our
age-structured McKendrick model. First, to allow for possible
population shifts or shifts in the onset of SUD, we include two
age-stratified influxes of new SUD cases in the shape of gamma
distributions peaked at ages $a^{\rm max}_1$ and $a^{\rm max}_2$ with
amplitudes $r_1$ and $r_2$, respectively. Large values of $r_1$
compared to $r_2$ imply that the influx of new SUD cases occurs mostly
through the distribution that is peaked at $a^{\rm max}_1$ and
vice-versa.  Second, to take into account non-overdose deaths among
drug users, the mortality rate we use for the population with SUD is given by an
age-stratified baseline value derived from the
Gompertz--Makeham--Siler mortality
model~\cite{gompertz1825,makeham1860law,siler1979competing,siler1983parameters,cohen2018gompertz}
to which an excess drug-induced mortality $\mu_{\rm d}$ is added.

As typical in data assimilation, at each forecasting step we use new
overdose fatality data to update the system state and estimates of
model quantities (such as the drug-induced mortality rate
$\hat{\mu}_{\rm d}$, the ages at which the rate of the onset of SUD is maximal, $\hat{a}_1^{\rm max}, \hat{a}_2^{\rm max}$, and the amplitudes
of the SUD influx distributions $\hat{r}_1, \hat{r}_2$) and use these
values for subsequent forecasts. Since CDC WONDER data is available
from 1999 until 2021, a forecast for drug-overdose fatalities in year
$Y$ is based on assimilated observational data between 1999 and $Y-1$.
Figures~\ref{fig:US_forecast}(c--e) display age-stratified
  overdose forecasts (solid red curves) and
  corresponding observational data (dashed black curves) for the years
  2001, 2011, and 2021.
  These are representative years selected from
the three phases outlined above. In all panels red shaded regions indicate $3\sigma$
intervals and ages are binned in 5 year intervals. The shown nationwide forecasts exhibit a remarkable
similarity to the actual observations; despite a significant rise in
overdose fatalities between 2019 and 2020, the EnKF forecast of
age-stratified overdose deaths remains in close agreement with the
reported number of fatalities.

\begin{figure}[t]
    \centering
    \includegraphics{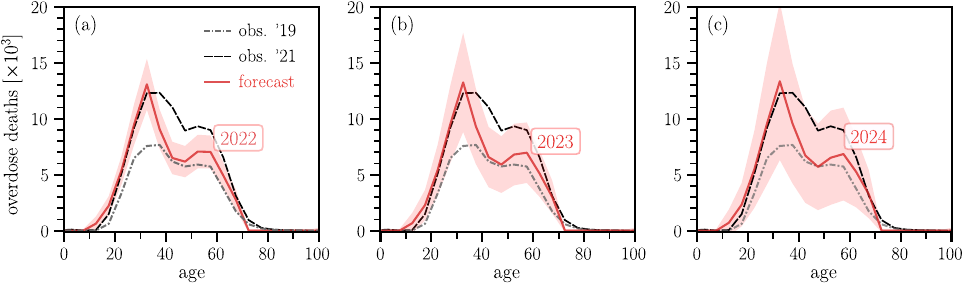}
    \caption{\textbf{Forecasting nationwide overdose fatalities for 2022--2024.}
      (a) Forecasts of overdose deaths in the United States as a
      function of age (0--100 years) for the year 2022. Data for 2022
      is not available at the time of writing (Summer 2023) and will
      be made public by the CDC in early 2024. The forecast (solid red
      curve) is higher than the 2019 observations (dash-dotted gray
      curve) for all age groups but less than the 2021 data (dashed
      black curve) for ages greater than 30. Light red shaded regions
      show values within 3$\sigma$. The predicted number of overdose
      deaths will stay near the record highs recorded in 2021 among
      younger persons, until age 30. (b,c) Forecasts of overdose
      deaths in the United States as a function of age (0--100 years)
      for the years 2023 and 2024. Predicted trends are similar to
      2022 projections and indicate that overdose fatalities
      for those under age 30 will remain high in the near future. As there
      is no available data for 2022 to 2024 the 3$\sigma$ intervals increase due to the larger uncertainty as time
      progresses. Ages are binned in 5 year intervals and prediction values are displayed at the center of each
        bin interval.}       
    \label{fig:future_forecast}
\end{figure}

In addition to forecasting fatalities in each age group, we also used our
EnKF to estimate the trajectories of $\hat{\mu}_{\rm d}$,
$\hat{a}^{\rm max}_1, \hat{a}^{\rm max}_2$ and $\hat{r}_1$ and
$\hat{r}_2$ over the 1999--2021 interval. The EnKF was initialized
with $\hat{\mu}_{\rm d} = 0.2 \%$ per year, consistent with 1999 data \cite{SAMHSI2010};
we also set the initial values $\hat{a}^{\rm max}_1 = 30$ years,
$\hat{a}^{\rm max}_2 = 45$ years and $\hat{r}_1 =\hat{r}_2= 2\%$ per
year. Figure~\ref{fig:US_forecast}(f) shows that $\hat{\mu}_{\rm d}$
increased more than 7-fold in the past 20 years, rising from about
$0.2 \%$ per year to $1.5 \%$ per year. This finding is largely
independent of the initial value of $\hat{\mu}_{\rm d}$; the final
estimate, $\hat{\mu}_{\rm d} = 1.5 \%$ overdose deaths per year is
larger than the baseline Gompertz--Makeham--Siler mortality rates for
all ages under 60 years old.

The trajectory of the quantities $\hat{a}^{\rm max}_1, \hat{a}^{\rm
  max}_2$ presented in Fig.~\ref{fig:US_forecast}(g) show that between
1999 and 2009 these values remain approximately stable within the
30--35 and 45--50 year range, respectively.  In later years however,
while $\hat{a}^{\rm max}_2$ decreases only mildly, there is a strong
descent of $\hat{a}^{\rm max}_1$ towards lower values, even below 20
years, indicating a substantial inflow of SUD cases at lower ages.
Figure~\ref{fig:US_forecast}(g) also shows that $\hat{r}_1$ increases
within the period of observation and that in 2015 it surpasses
$\hat{r}_2$, so that the influx of persons with SUD is dominated by
the distribution peaked at the younger age $\hat{a}^{\rm max}_1$. The
shift of the onset of SUD towards younger ages that is observed
starting in 2015 is consistent with the concurrent emergence of high
mortality rates within the 25--35 age group as seen in
Fig.~\ref{fig:US_forecast}(b).

Assessing the mortality rate of the population with SUD is challenging since a
large number of subjects must be recruited and followed to evaluate
the occurrence of a rare event such as death.  Typical studies are
conducted among SUD patients enrolled in treatment clinical trials or
who have been hospitalized and monitored post-discharge. These studies
reveal that drug users exhibit elevated mortality compared to the
general population primarily due to fatal overdoses, but also due to
viral infections, cardiovascular disease, and cancer
\cite{Mathers2013, Lindblad2016}.  Overdose-specific mortality rates
among SUD patients vary between 0.2$\%$ and 1$\%$ depending on gender,
drug of choice, treatment type \cite{Lindblad2016, Hser2017,
  Sordo2017, King2022}; other meta-studies reveal that the mortality
rate for persons who abuse opioids is about 0.7$\%$
\cite{Bahji2020}. Our estimates for $\hat{\mu}_{\rm d}$ from
Fig.~\ref{fig:US_forecast}(e) are in agreement with these values; in
addition, our work enables the tracking of longitudinal changes in
overdose mortality rates throughout the entire 1999--2021 period,
uncovering a significant and alarming surge in mortality rates among
individuals with SUD. Particularly noteworthy is the pronounced
increase observed during the years 2020 and 2021.

Finally, in Fig.\,\ref{fig:future_forecast} we show nationwide
forecasts for the years 2022, 2023, 2024 using our EnKF and the 2021
values of $\hat{\mu}_d, \hat{a}_1^{\rm max}, \hat{a}_2^{\rm max},
\hat{r}_1, \hat{r}_2$. These are overdose fatality forecasts for which observational data are not available at the time of writing (Summer 2023); data for the year 2022 will only be made public by the CDC in early 2024. Our forecasts show that
the number of overdose fatalities in the next few years will remain
high and will continue to exceed those reported in 2019 for all age
groups. Fatalities will remain largest among those younger than 30,
with drug overdose counts matching the record values of the pandemic
year 2021. However, we expect to see a decrease in
overdose deaths in age groups older than 30.

To summarize, our findings reveal a staggering 7-fold increase in
mortality rates among individuals with SUD, a generational shift
towards drug use at younger ages, and alarming numbers of past and
projected overdose deaths among individuals up to 30 years old. These
results emphasize the necessity for more focused approaches in
intervention and prevention strategies.
\subsection*{County-level variation}
\begin{figure}[htp!]
    \centering
    \includegraphics{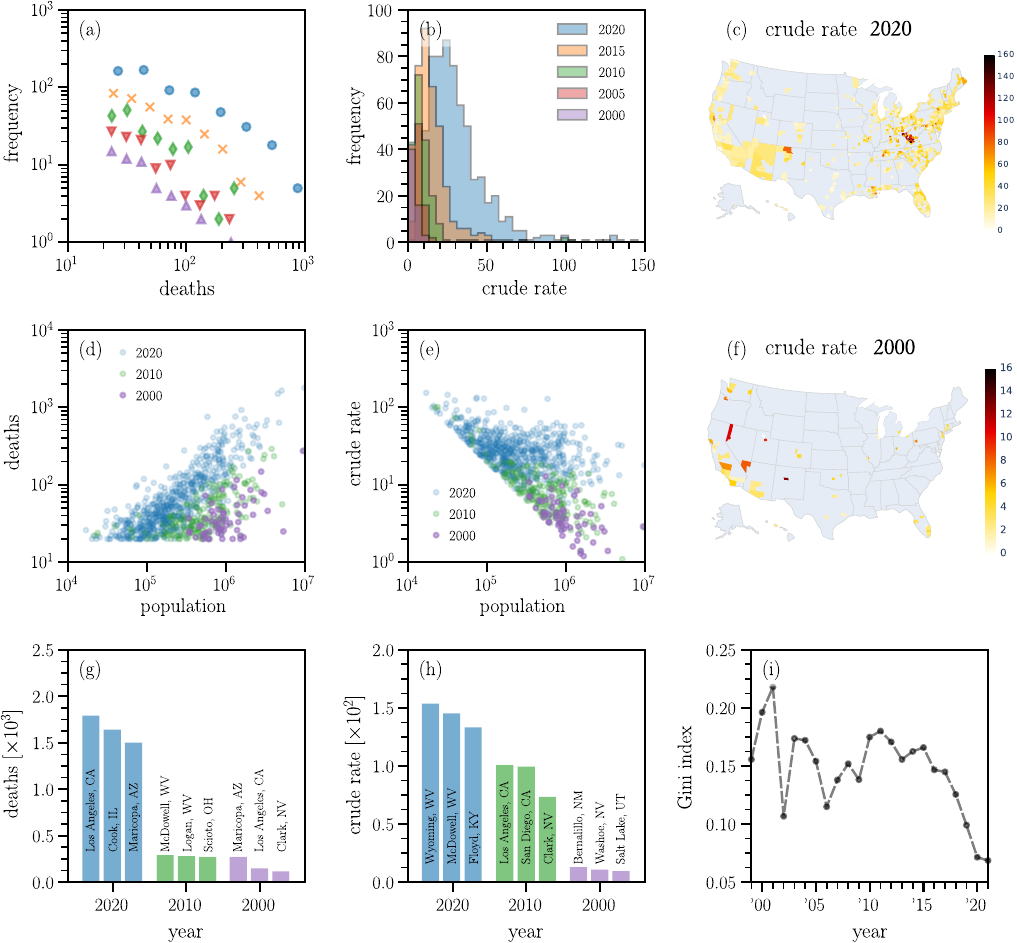}
    \caption{\textbf{Drug-overdose fatalities in the United States at
        the county level.} (a) Histogram of county-stratified
      drug-overdose fatalities for different years (blue disks: 2020,
      orange crosses: 2015, green diamonds: 2010, red inverted
      triangles: 2005, purple triangles: 2000). (b) Histogram of
      county-stratified crude rate data for different years. (d)
      County-stratified drug-overdose fatalities as a function of the
      corresponding county populations. (e) County-stratified crude
      rates as a function of the corresponding county
      populations. (c,f) Crude rates across different counties in 2020
      and 2000. The scales of the two color bars differ by a factor of
      10. In the gray regions, either no data or a statistically not
      significant number of cases were reported. In all panels,
      we did not include data for which at least one database entry
      (\eg, fatalities and crude rate) was marked unreliable. The
      minimum number of deaths in the remaining data is 20. Hence, the
      crude rate data in (e) lies above of the minimum crude rate given by 
      $2\times 10^6/\mathrm{Population}$. (g) The three counties with
      the largest overdose death tolls in 2020, 2010, and 2000. (h)
      The three counties with the largest overdose crude rates in
      2020, 2010, and 2000. (i) The Gini index across different
      years. A Gini index of 0 means that the crude rate is the same
      across all counties. If all overdose fatalities were
      concentrated in one out of $N_{\rm c}$ counties, the Gini index
      would be $1-1/N_{\rm c}$. The number of counties with
      statistically significant fatality counts (greater than 10
      deaths in a given year) and crude rates are $N_{\rm c} = 61$ in
      1999 and $N_{\rm c} = 742$ in 2021.}
        \label{fig:county_data}
\end{figure}
Although the overall number of drug-overdose fatalities in the United
States is rising, it varies significantly across
counties. Figure~\ref{fig:county_data}(a) shows the distribution of
county-stratified drug-overdose fatalities for select years between
2000--2020. Only counties with statistically significant fatalities of
more than 10 individuals per year are shown. The number of counties that
reached this significance threshold increased
from 61 counties (out of $3,147$) in 1999 to 742 (out of $3,142$) in
2021, as reported in the CDC WONDER database.  Between 1999 to 2021,
numerous counties reported annual numbers of overdose fatalities below
100. However, during the same period, the number of counties
experiencing between 100 and 1,000 annual overdose fatalities steadily
increased. In 2020 (blue disks) a few counties even recorded close to
$1,000$ overdose deaths. Crude rates, defined as the number of deaths
per $100,000$ persons, also increased significantly over time, as seen
in Fig.~\ref{fig:county_data}(b): in the year 2000 the mean crude rate
among all counties for which data was available was 4.3 cases per
$100,000$, in 2020 it was 31.5 cases per 100,000. This 7-fold increase
is consistent with the similar rise in $\hat{\mu}_d$ as inferred by
our EnKF on the national level. The distributions in
Fig.~\ref{fig:county_data}(b) also show that the crude rates exhibit a
high degree of variability across counties.

Figure~\ref{fig:county_data}(d) shows that in 2000, only counties with
populations larger than $100,000$ residents experienced statistically
significant numbers of drug-overdose fatalities.  In the years since,
crude rates substantially increased for these counties, especially
between 2010 and 2020, while smaller counties with population sizes of
about $10,000$ also started reporting significant numbers of fatal
overdoses, as shown in Fig.~\ref{fig:county_data}(e).  This indicates
that the drug overdose epidemic has permeated all jurisdictions,
regardless of population.

The heat maps in Figs.~\ref{fig:county_data}(c,f) confirm the increase
in the number of counties affected by the drug epidemic between 2000
and 2020. Notice that the scales of the two color bars differ by a
factor of 10.  Figure~\ref{fig:county_data}(c) shows that in 2000, the
most affected areas were population centers in the Western United
States: that year, the largest overdose fatality counts occurred in
Maricopa County, AZ; Los Angeles County, CA; and Clark County, NV
[Fig.~\ref{fig:county_data}(g)] and the largest crude rates were
reported in Bernalillo County, NM; Washoe County, NV and Salt Lake
County, UT [Fig.~\ref{fig:county_data}(h)]. In 2020, many regions in
the Central and Eastern United States also became heavily impacted by
the drug epidemic, including many smaller population counties
[Fig.~\ref{fig:county_data}(f)]. The largest 2020 fatality county were
reported in Los Angeles County, CA; Cook County, IL; and Maricopa
County, AZ [Fig.~\ref{fig:county_data}(g)]. The largest 2020 crude
rates were registered in Wyoming County, WV; McDowell County, WV; and
Floyd County, KY [Fig.~\ref{fig:county_data}(h)].

In 2000, Los Angeles County, the most populous in the United States,
accounted for approximately 12\% of the total population of the 59
counties with statistically significant overdose fatalities. Its
proportion of overdose fatalities was about 10\%. However, due to more
counties reporting large numbers of overdose deaths, by 2020, Los
Angeles County's population represented only 4\% of the total
population of the 640 counties with statistically significant overdose
deaths.  That same year, Los Angeles County contributed to
approximately 3\% of the registered overdose deaths at the county
level.  Furthermore, in 2020, despite the 10 most populous counties in
the United States being home to roughly 16\% of the population (out of
the 258 million associated with the 640 counties with statistically
significant overdose fatalities), they recorded less than 13\% of the
number of overdose fatalities among these 640 counties.  Conversely,
the 50 least populous counties, with less than 1\% of the total
population among the 640 affected counties, accounted for more than
2\% of the number of overdose deaths. These statistics highlight the
shifting patterns of overdose fatalities, with a notable rise of
overdose deaths in less populated counties.
To better understand how the drug deaths are shared across
counties, we consider the Gini
coefficient~\cite{gini1912variabilita,xu2020diversity}, a measure of
inequality among a set of $N_{\rm c}$ values of a distribution; in
our case $N_{\rm c}$ is the number of counties that have reported
statistically significant numbers of overdose deaths.
We compute the Gini index by plotting the proportion of the
total number of overdose fatalities accumulated across counties
against the cumulative population fraction across counties. The lower
bound for the Gini index is 0 (perfect equality, indicating
that overdose deaths and county populations are proportional), and
the upper bound is $1 - 1/N_{\rm c}$ (perfect inequality, indicating
that all overdose deaths occurred within a single county). We find
that the Gini coefficient dropped from a value of about 0.2 in the
year 2000 $(N_{\rm c} = 59)$, to about 0.07 in the year 2021 ($N_{\rm
  c} = 640$), which is consistent with increases in the number of
counties affected by the drug-overdose epidemic.

In the next section, we employ the modeling and forecasting techniques
established in the previous section to examine the progression of
age-specific overdose fatality counts in three specific regions: Los
Angeles County, CA; Cook County, IL; and the combined area of New York
City's five boroughs (The Bronx, Brooklyn, Manhattan, Queens, and Staten
Island).
\subsection*{Forecasting overdose fatalities in three counties}
\begin{figure}
    \centering
    \includegraphics{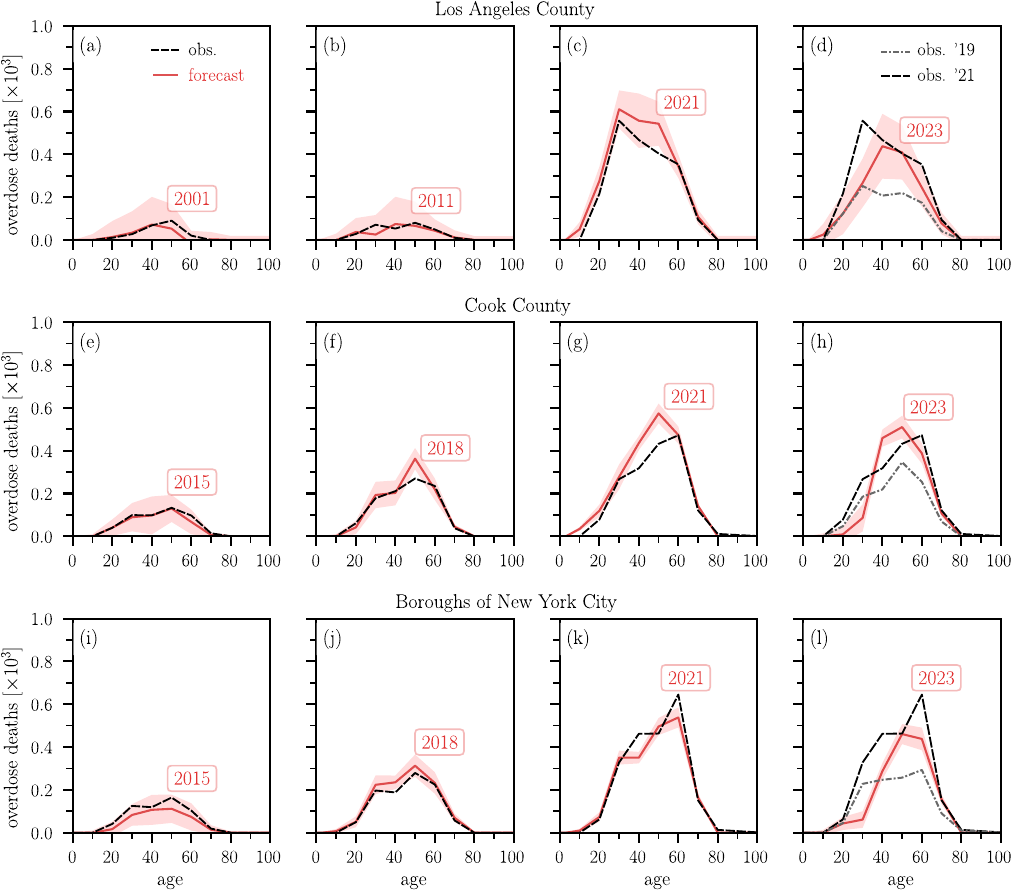}
    \caption{\textbf{Forecasting overdose fatalities in select US
        jurisdictions.} (a--d) Forecasts of overdose deaths in Los
      Angeles County as a function of age (0--100 years) for 2001,
      2011, 2021, and 2023. Solid red curves and shaded regions
      indicate mean predictions and 3$\sigma$ intervals,
      respectively. Observed fatalities are indicated by dashed black
      curves. Because observational data for 2023 is not available at
      the time of writing, we show 2019 data (dash-dotted gray curve)
      and 2021 data (dashed black curve) in panels (d,h,l). (e--h)
      Forecasts for Cook County using the same graphic
      representations as in panels (a--d) and for the years 2015,
      2018, 2021, 2023. (i--l) Forecasts for the five boroughs that
      comprise New York City (The Bronx, Brooklyn, Manhattan, Queens, and
      Staten Island) using the same graphic representations as in
      panels (a--d) and for the years 2007, 2014, 2021, 2023. Although
      the numerical escalation in drug-overdose deaths in Cook County
      (2015--2021) and New York City (2007--2021) is similar to what
      is observed for Los Angeles County, the timelines are much
      accelerated. In 2021 the population of Los Angeles County was
      9.7 million, in Cook County 5.1 million and in New York City 8.3
      million, indicating a more acute crisis in Cook
      County. Ages are binned in 10 year intervals and prediction values are displayed at the center of each bin interval.}
    \label{fig:county_forecast}
\end{figure}
The dynamics of overdose fatalities in Los Angeles County, Cook
County and five boroughs of New York City have unfolded in
substantially different ways over the past two decades. In 1999, Cook
County reported a total of 14 overdose fatalities, New York City
recorded 52, and Los Angeles County 384. By 2021, these numbers had
risen to 1,699 overdose deaths for Cook County, 2,091 for Los Angeles
County, and 2,124 for New York City.  Cook County experienced the most
striking rise in overdose fatalities between 1999 to 2021: an
unprecedented 120-fold increase.  In comparison, New York City
experienced a 40-fold increase and Los Angeles County a 5-fold
increase.  These numbers are even more striking given Cook County's
smaller population (5.1 million in 2021), compared to the population of Los Angeles
County (9.7 million in 2021) and New York City (8.3 million in 2021).

As in our nationwide forecasts, we use an EnKF in conjunction with an
age-structured overdose mortality model that accounts for the underlying age variation in the county populations.
Because of the relatively small overdose fatality counts in Cook
County and New York City in the early 2000s, we do not report
forecasts for 2001 as done nationwide and for Los Angeles County,
but we use years from 2013 onwards for which enough data is available
across all age groups. Specifically, in Cook County, the total number
of reported age-stratified drug-overdose fatalities remained below 100
for most years prior to 2013.  This points to a delayed emergence of
the overdose death crisis in this county which became extraordinarily
acute in just a few years. A finer analysis reveals that the largest
increases in drug overdose mortality in Cook County are due to heroin
(a 10-fold rise between 2012 and 2013) and to fentanyl (a 5-fold rise
between 2015 and 2016).

Results are shown in Fig.~\ref{fig:county_forecast}. The first three
historical forecasts for each region are compared to the corresponding
observational data (dashed black curves). Since the observational data for
2023 is not available at the time of writing (Summer 2023), we display
the 2019 data (dash-dotted gray curve) and the 2021 data (dashed black
curve) in the 2023 forecast. For Los Angeles County, we initialized
the EnKF with $\hat{\mu}_{\rm d} = 0.25 \%$ per year and $\hat{r}_1
=\hat{r}_2=2\%$ per year. The simulations for Cook County and
New York City start in 2013, and we initially set $\hat{\mu}_{\rm d} = 0.5
\%$ per year and $\hat{r}_1 =\hat{r}_2=6\%$ per year. The estimated
drug-caused mortality rate for Cook County and New York City in 2021
are, respectively, $\hat{\mu}_{\rm d}=1.3\%$ per year and
$\hat{\mu}_{\rm d}=1.1\%$, substantially larger than the 2021
estimates for Los Angeles County ($\hat{\mu}_{\rm d}=0.7\%$ per
year). This is also consistent with the much larger increase in
overdose fatalities in Cook County and New York City compared to Los
Angeles County.
\section*{Discussion}
Escalating drug-induced deaths have been a major public-health
challenge in the United States for more than a century. The
 over-prescription of morphine and opium led to an epidemic that affected almost 5 in 1,000 Americans in the 1890s~\cite{Courtwright2001}. This widespread crisis spurred a number of acts and regulations in the early 20$^{\rm th}$ century
 that succeeded in lowering opiate use and mortality rates~\cite{Musto1999, Courtwright1992}.
 The current epidemic involves a significantly higher prevalence of SUD
 and has unfolded via distinct spatio-temporal mortality waves driven by different
 drug types and localized sub-epidemics. Being able to forecast the complex evolution of fatal drug overdoses at the national, regional or county levels, would represent major advancements in helping curb drug abuse. 

In this work, we developed a forecasting method that combines an
age-structured model of addiction and overdose mortality with
observational data derived from the CDC WONDER database through a data
assimilation approach. By applying our method to nationwide data as
well as to three representative areas (Los Angeles County, Cook
County, and the five boroughs of New York City) we showed its ability
to provide near-term forecasts, to extract epidemiological
parameters, and to capture the heterogeneity in overdose mortality
across different counties. Since the demographics and geography of drug abuse are in constant flux, 
we believe our data assimilation approach holds promise for
informing targeted prevention and preparedness interventions aimed at
curbing drug overdose deaths.

The nationwide drug-induced mortality rate has risen more than 7-fold
in the past two decades, surpassing 1.5\% among persons suffering from
SUD. This rate exceeds the baseline Gompertz--Makeham--Siler mortality rates for all groups under 60 years old.  Our county level analysis reveals significant variations in overdose fatality trends.  For example, although at the onset of
1999--2021 period, Los Angeles County had a higher drug-induced
mortality rate than Cook County and New York City, in 2021 it had the
lowest, at 0.7$\%$, compared to Cook County at 1.3$\%$ and New York
City at 1.1$\%$. Furthermore, the annual number of overdose deaths in
Cook County and New York City grew much faster than in Los Angeles
County in recent years.  This points to a delayed, yet severe growth
of drug overdoses in New York City and especially in Cook County. For
past years, our predictions are in good agreement with tallied data.
For the year 2023 we predict slight decreases of drug overdoses
compared to the pandemic year 2021, both nationwide and in the
counties we surveyed. Specifically, we expect drug overdose deaths to
decrease nationwide by 17$\%$ compared to the values recorded in 2021,
by 24$\%$ in Los Angeles County, by 32$\%$ in New York City but only
by 8$\%$ in Cook County.


We also find that the overdose epidemic has spread to more counties
over time. In the year 1999, 61 counties out of $3,147$ had
statistically significant overdose fatalities, whereas this number
rose to 742 out of $3,142$ in 2021.  Not only has the number of
affected counties grown, but their relative contributions to the
overall overdose fatality count have become more evenly distributed
over the years. This finding implies that managing the overdose
epidemic cannot be simply accomplished by targeting a few specific
counties, rather each jurisdiction must develop specific plans
tailored to their unique socio-demographic and economic profiles.

Several limitations of this study are noteworthy.  Our findings are
based on four drug categories with the highest crude rates available
in the CDC WONDER database: fentanyl (T40.4), prescription opioids
(T40.2), heroin (T40.1), and methamphetamines (T43.6). We did not
include other categories such as T40.3 (methadone) or T40.5 (cocaine)
in our analysis due to their lower mortality rates. The dynamics of
fatalities associated with these drug categories may differ from the
fatality trends observed in our analysis.  Furthermore, in certain
jurisdictions, fatality data is unavailable as the CDC WONDER portal suppresses
entries where the number of deaths is less than 10. Additionally, some
overdose cases may involve multiple drugs. In such instances, deaths
are counted in all relevant categories, resulting in multiple
counts. Finally, comparisons of opioid-related overdose death rates at
the national, state, and county levels may be influenced by
significant variations in the reporting of specific drugs involved in
overdose deaths. Changes in drug reporting specificity over time and
across different states and counties can lead to potentially
misleading conclusions regarding actual drug-specific death
rates~\cite{jones2019data}.

There are several potential avenues for future work.  While we perform
yearly parameter updates and forecasts using final data from CDC
WONDER, it is also possible to utilize provisional fatality data,
which are released at earlier dates and on a monthly basis, to
generate more timely but similarly provisional forecasts;
retrospective updates can then be performed once final data 
becomes available. Although we only analyzed three large urban areas,
our method can be applied to other jurisdictions. In less populous
areas the number of fatalities may not be sufficiently large for a
meaningful age-stratified analysis; in these cases pooling data from
several neighboring jurisdictions with similar socio-economic
characteristics, using larger age-binning or considering biannual
forecasts may yield more meaningful results. Alternatively, for
small-number cases, a stochastic version of the
 Kermack--McKendrick model may be used as to evolve the state
 variables probabilistically \cite{chou2016hierarchical}.  Our
methods can also be used to forecast regional drug-overdose mortality
by gender, race or drug-type; the resulting projections for specific
groups of users or drugs of abuse may help guide more effective
intervention efforts. Large deviations of observed data from our
projections would signal fundamental changes to the illicit drug
landscape in the form of effective prevention and treatment programs,
or in the consumption of more addictive or lethal substances.
Finally, one may also study the numerical stability and forecasting
accuracy alternative ensemble-based Kalman filters, such as ensemble
adjustment Kalman filters~\cite{anderson2001ensemble}, or incorporate
backward passes and smoothing techniques into our method to
potentially enhance earlier parameter
estimates~\cite{evensen2000ensemble}.

\section*{Acknowledgements}
TC and MRD acknowledge financial support from the ARO through grant W911NF-18-1-0345; 
LB and MRD acknowledge financial support from the ARO through grant W911NF-23-1-0129.
\section*{Competing interests}
The authors declare no competing interests.
\section*{Data and code availability}
Our source codes are publicly available at
\url{https://gitlab.com/ComputationalScience/overdose-da}. The CDC
WONDER portal can be accessed at \url{https://wonder.cdc.gov/}.
\bibliographystyle{naturemag}
\bibliography{refs}

\begin{thebibliography}{10}
\expandafter\ifx\csname url\endcsname\relax
  \def\url#1{\texttt{#1}}\fi
\expandafter\ifx\csname urlprefix\endcsname\relax\def\urlprefix{URL }\fi
\providecommand{\bibinfo}[2]{#2}
\providecommand{\eprint}[2][]{\url{#2}}

\bibitem{mattson2021trends}
\bibinfo{author}{Mattson, C.~L.} \emph{et~al.}
\newblock \bibinfo{title}{Trends and geographic patterns in drug and synthetic
  opioid overdose deaths -- {U}nited {S}tates, 2013--2019}.
\newblock \emph{\bibinfo{journal}{Morbidity and Mortality Weekly Reports}}
  \textbf{\bibinfo{volume}{70}}, \bibinfo{pages}{202--207}
  (\bibinfo{year}{2021}).

\bibitem{ODonnell2020}
\bibinfo{author}{O'Donnell, J.}, \bibinfo{author}{Gladden, R.~M.},
  \bibinfo{author}{Mattson, C.~L.}, \bibinfo{author}{Hunter, C.} \&
  \bibinfo{author}{Davis, N.~L.}
\newblock \bibinfo{title}{Vital signs: {C}haracteristics of drug overdose
  deaths involving opioids and stimulants - 24 states and the {District of
  Columbia}, {January -- June} 2019}.
\newblock \emph{\bibinfo{journal}{Morbidity and Mortality Weekly Reports}}
  \textbf{\bibinfo{volume}{69}}, \bibinfo{pages}{1189--1197}
  (\bibinfo{year}{2020}).

\bibitem{Jones2018}
\bibinfo{author}{Jones, C.~M.}, \bibinfo{author}{Einstein, E.~B.} \&
  \bibinfo{author}{Compton, W.~M.}
\newblock \bibinfo{title}{{Changes in synthetic opioid involvement in drug
  overdose deaths in the United States, 2010-2016}}.
\newblock \emph{\bibinfo{journal}{JAMA}} \textbf{\bibinfo{volume}{319}},
  \bibinfo{pages}{1819--1821} (\bibinfo{year}{2018}).

\bibitem{Armenian2018}
\bibinfo{author}{Armenian, P.}, \bibinfo{author}{Vo, K.~T.},
  \bibinfo{author}{Barr-Walker, J.} \& \bibinfo{author}{Lynch, K.~L.}
\newblock \bibinfo{title}{Fentanyl, fentanyl analogs and novel synthetic
  opioids: {A} comprehensive review}.
\newblock \emph{\bibinfo{journal}{Neuropharmacology}}
  \textbf{\bibinfo{volume}{134}}, \bibinfo{pages}{121--132}
  (\bibinfo{year}{2018}).

\bibitem{Forman2006}
\bibinfo{author}{Forman, R.~F.} \& \bibinfo{author}{Block, L.~G.}
\newblock \bibinfo{title}{The marketing of opioid medications without
  prescription over the internet}.
\newblock \emph{\bibinfo{journal}{Journal of Public Policy and Marketing}}
  \textbf{\bibinfo{volume}{25}}, \bibinfo{pages}{133--146}
  (\bibinfo{year}{2006}).

\bibitem{Mackey2017}
\bibinfo{author}{Mackey, T.~K.}, \bibinfo{author}{Kalyanam, J.},
  \bibinfo{author}{Katsuki, T.} \& \bibinfo{author}{Lanckriet, G.}
\newblock \bibinfo{title}{Twitter-based detection of illegal online sale of
  prescription opioid}.
\newblock \emph{\bibinfo{journal}{American Journal of Public Health}}
  \textbf{\bibinfo{volume}{107}}, \bibinfo{pages}{1910--1915}
  (\bibinfo{year}{2017}).

\bibitem{Lamy2020}
\bibinfo{author}{Lamy, F.~R.} \emph{et~al.}
\newblock \bibinfo{title}{Listed for sale: {A}nalyzing data on fentanyl,
  fentanyl analogs and other novel synthetic opioids on one cryptomarket}.
\newblock \emph{\bibinfo{journal}{Drug and Alcohol Dependence}}
  \textbf{\bibinfo{volume}{213}}, \bibinfo{pages}{108115}
  (\bibinfo{year}{2020}).

\bibitem{Duhart2022}
\bibinfo{author}{{Duhart-Clarke}, S.~E.}, \bibinfo{author}{Kral, A.~H.} \&
  \bibinfo{author}{Zibbell, J.~E.}
\newblock \bibinfo{title}{Consuming illicit opioids during a drug overdose
  epidemic: {Illicit} fentanyls, drug discernment, and the radical
  transformation of the illicit opioid market}.
\newblock \emph{\bibinfo{journal}{International Journal of Drug Policy}}
  \textbf{\bibinfo{volume}{99}}, \bibinfo{pages}{103467}
  (\bibinfo{year}{2022}).

\bibitem{Kariisa2023}
\bibinfo{author}{Kariisa, M.}, \bibinfo{author}{O'Donnell, J.},
  \bibinfo{author}{Kumar, S.}, \bibinfo{author}{Mattson, C.~L.} \&
  \bibinfo{author}{Goldberger, B.~A.}
\newblock \bibinfo{title}{Illicitly manufactured fentanyl involved overdose
  deaths with detected xylazine -- {United States}, {January 2019 -- June
  2022}}.
\newblock \emph{\bibinfo{journal}{Morbidity and Mortality Weekly Reports}}
  \textbf{\bibinfo{volume}{72}}, \bibinfo{pages}{721--727}
  (\bibinfo{year}{2023}).

\bibitem{Stein2017}
\bibinfo{author}{Stein, E.~M.}, \bibinfo{author}{Gennuso, K.~P.},
  \bibinfo{author}{Ugboaja, D.~C.} \& \bibinfo{author}{Remington, P.~L.}
\newblock \bibinfo{title}{The epidemic of despair among {White Americans:
  Trends} in the leading causes of premature death, 1999-2015}.
\newblock \emph{\bibinfo{journal}{American Journal of Public Health}}
  \textbf{\bibinfo{volume}{107}}, \bibinfo{pages}{1541--1547}
  (\bibinfo{year}{2017}).

\bibitem{Case2020}
\bibinfo{author}{Case, A.} \& \bibinfo{author}{Deaton, A.}
\newblock \emph{\bibinfo{title}{Deaths of despair and the future of
  capitalism}} (\bibinfo{publisher}{Princeton University Press},
  \bibinfo{address}{Princeton, NJ}, \bibinfo{year}{2020}).

\bibitem{Friedman2021}
\bibinfo{author}{Friedman, J.} \& \bibinfo{author}{Akre, S.}
\newblock \bibinfo{title}{{COVID-19 and the drug overdose crisis: Uncovering
  the deadliest months in the United States, January -- July 2020}}.
\newblock \emph{\bibinfo{journal}{American Journal of Public Health}}
  \textbf{\bibinfo{volume}{111}}, \bibinfo{pages}{1284--1291}
  (\bibinfo{year}{2021}).

\bibitem{dorsogna2023fentanyl}
\bibinfo{author}{D'Orsogna, M.~R.}, \bibinfo{author}{B{\"o}ttcher, L.} \&
  \bibinfo{author}{Chou, T.}
\newblock \bibinfo{title}{Fentanyl-driven acceleration of racial, gender and
  geographical disparities in drug overdose deaths in the {U}nited {S}tates}.
\newblock \emph{\bibinfo{journal}{PLOS Global Public Health}}
  \textbf{\bibinfo{volume}{3}}, \bibinfo{pages}{e0000769}
  (\bibinfo{year}{2023}).

\bibitem{Jalal2018}
\bibinfo{author}{Jalal, H.} \emph{et~al.}
\newblock \bibinfo{title}{Changing dynamics of the drug overdose epidemic in
  the {United States} from 1979 through 2016}.
\newblock \emph{\bibinfo{journal}{Science}} \textbf{\bibinfo{volume}{361}},
  \bibinfo{pages}{6408} (\bibinfo{year}{2018}).

\bibitem{Peters2020}
\bibinfo{author}{Peters, D.~J.}, \bibinfo{author}{Monnat, S.~M.},
  \bibinfo{author}{Hochstetler, A.~L.} \& \bibinfo{author}{Berg, M.~T.}
\newblock \bibinfo{title}{The opioid hydra: {U}nderstanding overdose mortality
  epidemics and syndemics across the rural-urban continuum}.
\newblock \emph{\bibinfo{journal}{Rural Sociology}}
  \textbf{\bibinfo{volume}{85}}, \bibinfo{pages}{589--622}
  (\bibinfo{year}{2020}).

\bibitem{powell2023trends}
\bibinfo{author}{Powell, D.}, \bibinfo{author}{Shetty, K.~D.} \&
  \bibinfo{author}{Peet, E.~D.}
\newblock \bibinfo{title}{Trends in overdose deaths involving gabapentinoids
  and {Z}-drugs in the {U}nited {S}tates}.
\newblock \emph{\bibinfo{journal}{Drug and Alcohol Dependence}}
  \textbf{\bibinfo{volume}{249}}, \bibinfo{pages}{109952}
  (\bibinfo{year}{2023}).

\bibitem{Segel2021}
\bibinfo{author}{Segel, J.~E.} \& \bibinfo{author}{Winkelman, T. N.~A.}
\newblock \bibinfo{title}{Persistence and pervasiveness: {E}arly wave opioid
  overdose death rates associated with subsequent overdose death rates}.
\newblock \emph{\bibinfo{journal}{Public Health Reports}}
  \textbf{\bibinfo{volume}{136}}, \bibinfo{pages}{212--218}
  (\bibinfo{year}{2021}).

\bibitem{Blanco2022}
\bibinfo{author}{Blanco, C.}, \bibinfo{author}{Wall, M.~M.} \&
  \bibinfo{author}{Olfson, M.}
\newblock \bibinfo{title}{Data needs and models for the opioid epidemic}.
\newblock \emph{\bibinfo{journal}{Molecular Psychiatry}}
  \textbf{\bibinfo{volume}{27}}, \bibinfo{pages}{787--792}
  (\bibinfo{year}{2022}).

\bibitem{Lim2022}
\bibinfo{author}{Lim, T.~Y.} \emph{et~al.}
\newblock \bibinfo{title}{Modeling the evolution of the {US} opioid crisis for
  national policy development}.
\newblock \emph{\bibinfo{journal}{Proceedings of the National Academy of
  Sciences of the United States of America}} \textbf{\bibinfo{volume}{119}},
  \bibinfo{pages}{e2115714119} (\bibinfo{year}{2022}).

\bibitem{Borquez2022}
\bibinfo{author}{Borquez, A.} \& \bibinfo{author}{Martin, N.~K.}
\newblock \bibinfo{title}{Fatal overdose: {P}redicting to prevent}.
\newblock \emph{\bibinfo{journal}{International Journal of Drug Policy}}
  \textbf{\bibinfo{volume}{104}}, \bibinfo{pages}{103677}
  (\bibinfo{year}{2022}).

\bibitem{Monnat2018}
\bibinfo{author}{Monnat, S.~M.}
\newblock \bibinfo{title}{Factors associated with county-level differences in
  {U.S.} drug-related mortality rates}.
\newblock \emph{\bibinfo{journal}{American Journal of Preventive Medicine}}
  \textbf{\bibinfo{volume}{54}}, \bibinfo{pages}{611--619}
  (\bibinfo{year}{2018}).

\bibitem{Rigg2018}
\bibinfo{author}{Rigg, K.}, \bibinfo{author}{Monnat, S.~M.} \&
  \bibinfo{author}{Chavez, M.~N.}
\newblock \bibinfo{title}{Opioid-related mortality in rural {A}merica:
  {G}eographic heterogeneity and intervention strategies}.
\newblock \emph{\bibinfo{journal}{International Journal of Drug Policy}}
  \textbf{\bibinfo{volume}{57}}, \bibinfo{pages}{119--129}
  (\bibinfo{year}{2018}).

\bibitem{Brownstein2010}
\bibinfo{author}{Brownstein, J.~S.}, \bibinfo{author}{Green, T.~C.},
  \bibinfo{author}{Cassidy, T.~A.} \& \bibinfo{author}{Butler, S.~F.}
\newblock \bibinfo{title}{Geographic information systems and
  pharmacoepidemiology: {U}sing spatial cluster detection to monitor local
  patterns of prescription opioid abuse}.
\newblock \emph{\bibinfo{journal}{Pharmacoepidemiological and Drug Safety}}
  \textbf{\bibinfo{volume}{19}}, \bibinfo{pages}{627--637}
  (\bibinfo{year}{2010}).

\bibitem{Basak2019}
\bibinfo{author}{Basak, A.}, \bibinfo{author}{Cadena, J.},
  \bibinfo{author}{Marathe, A.} \& \bibinfo{author}{Vullikanti, A.}
\newblock \bibinfo{title}{Detection of spatiotemporal prescription opioid hot
  spots with network scan statistics: {M}ultistate analysis}.
\newblock \emph{\bibinfo{journal}{JMIR Public Health and Surveillance}}
  \textbf{\bibinfo{volume}{5}}, \bibinfo{pages}{e12110} (\bibinfo{year}{2019}).

\bibitem{Campo2020}
\bibinfo{author}{Campo, D.~S.}, \bibinfo{author}{Gussler, J.~W.},
  \bibinfo{author}{Sue, A.}, \bibinfo{author}{Skums, P.} \&
  \bibinfo{author}{Khudyakov, Y.}
\newblock \bibinfo{title}{Accurate spatiotemporal mapping of drug overdose
  deaths by machine learning of drug-related web-searches}.
\newblock \emph{\bibinfo{journal}{PLOS One}} \textbf{\bibinfo{volume}{15}},
  \bibinfo{pages}{e0243622} (\bibinfo{year}{2020}).

\bibitem{Marks2021}
\bibinfo{author}{Marks, C.} \emph{et~al.}
\newblock \bibinfo{title}{Identifying counties at risk of high overdose
  mortality burden during the emerging fentanyl epidemic in the {USA}: {A}
  predictive statistical modelling study.}
\newblock \emph{\bibinfo{journal}{Lancet Public Health}}
  \textbf{\bibinfo{volume}{6}}, \bibinfo{pages}{e720--e728}
  (\bibinfo{year}{2021}).

\bibitem{Marks2021b}
\bibinfo{author}{Marks, C.} \emph{et~al.}
\newblock \bibinfo{title}{Methodological approaches for the prediction of
  opioid use-related epidemics in the {United States}: {A} narrative review and
  cross-disciplinary call to action}.
\newblock \emph{\bibinfo{journal}{Translational Research}}
  \textbf{\bibinfo{volume}{234}}, \bibinfo{pages}{88--113}
  (\bibinfo{year}{2021}).

\bibitem{Sumetsky2021}
\bibinfo{author}{Sumetsky, N.} \emph{et~al.}
\newblock \bibinfo{title}{Predicting the future course of opioid overdose
  mortality: {A}n example from two {U.S.} states}.
\newblock \emph{\bibinfo{journal}{Epidemiology}} \bibinfo{pages}{61--69}
  (\bibinfo{year}{2021}).

\bibitem{Wagner2021}
\bibinfo{author}{Wagner, N.~M.} \emph{et~al.}
\newblock \bibinfo{title}{Development and validation of a prediction model for
  opioid use disorder among youth}.
\newblock \emph{\bibinfo{journal}{Drug and Alcohol Dependence}}
  \textbf{\bibinfo{volume}{227}}, \bibinfo{pages}{108980}
  (\bibinfo{year}{2021}).

\bibitem{Stringfellow2022}
\bibinfo{author}{Stringfellow, E.~J.} \emph{et~al.}
\newblock \bibinfo{title}{Reducing opioid use disorder and overdose deaths in
  the united states: {A} dynamic modeling analysis}.
\newblock \emph{\bibinfo{journal}{Scientific Advances}}
  \textbf{\bibinfo{volume}{8}}, \bibinfo{pages}{eabm8147}
  (\bibinfo{year}{2022}).

\bibitem{crassidis2004optimal}
\bibinfo{author}{Crassidis, J.~L.} \& \bibinfo{author}{Junkins, J.~L.}
\newblock \emph{\bibinfo{title}{Optimal estimation of dynamic systems}}
  (\bibinfo{publisher}{Chapman and Hall/CRC}, \bibinfo{address}{Boca Raton,
  FL}, \bibinfo{year}{2004}).

\bibitem{law2015data}
\bibinfo{author}{Law, K.}, \bibinfo{author}{Stuart, A.} \&
  \bibinfo{author}{Zygalakis, K.}
\newblock \emph{\bibinfo{title}{{D}ata assimilation: {A} mathematical
  introduction}}.
\newblock Texts in Applied Mathematics (\bibinfo{publisher}{Springer},
  \bibinfo{address}{Cham, Switzerland}, \bibinfo{year}{2015}).

\bibitem{bottcher2023forecasting}
\bibinfo{author}{B\"ottcher, L.}, \bibinfo{author}{Chou, T.} \&
  \bibinfo{author}{D'Orsogna, M.~R.}
\newblock \bibinfo{title}{Modeling and forecasting age-specific overdose
  mortality in the {U}nited {S}tates}.
\newblock \emph{\bibinfo{journal}{European Physics Journal Special Topics}}
  (\bibinfo{year}{2023}).
\newblock \urlprefix\url{https://doi.org/10.1140/epjs/s11734-023-00801-z}.

\bibitem{bauer2015quiet}
\bibinfo{author}{Bauer, P.}, \bibinfo{author}{Thorpe, A.} \&
  \bibinfo{author}{Brunet, G.}
\newblock \bibinfo{title}{The quiet revolution of numerical weather
  prediction}.
\newblock \emph{\bibinfo{journal}{Nature}} \textbf{\bibinfo{volume}{525}},
  \bibinfo{pages}{47--55} (\bibinfo{year}{2015}).

\bibitem{lillacci2010parameter}
\bibinfo{author}{Lillacci, G.} \& \bibinfo{author}{Khammash, M.}
\newblock \bibinfo{title}{Parameter estimation and model selection in
  computational biology}.
\newblock \emph{\bibinfo{journal}{PLOS Computational Biology}}
  \textbf{\bibinfo{volume}{6}}, \bibinfo{pages}{e1000696}
  (\bibinfo{year}{2010}).

\bibitem{schneider2021epidemic}
\bibinfo{author}{Schneider, T.} \emph{et~al.}
\newblock \bibinfo{title}{Epidemic management and control through
  risk-dependent individual contact interventions}.
\newblock \emph{\bibinfo{journal}{PLOS Computational Biology}}
  \textbf{\bibinfo{volume}{18}}, \bibinfo{pages}{e1010171}
  (\bibinfo{year}{2022}).

\bibitem{pei2021identifying}
\bibinfo{author}{Pei, S.}, \bibinfo{author}{Liljeros, F.} \&
  \bibinfo{author}{Shaman, J.}
\newblock \bibinfo{title}{Identifying asymptomatic spreaders of
  antimicrobial-resistant pathogens in hospital settings}.
\newblock \emph{\bibinfo{journal}{Proceedings of the National Academy of
  Sciences of the United States of America}} \textbf{\bibinfo{volume}{118}},
  \bibinfo{pages}{e2111190118} (\bibinfo{year}{2021}).

\bibitem{li2020substantial}
\bibinfo{author}{Li, R.} \emph{et~al.}
\newblock \bibinfo{title}{Substantial undocumented infection facilitates the
  rapid dissemination of novel coronavirus ({SARS-CoV-2})}.
\newblock \emph{\bibinfo{journal}{Science}} \textbf{\bibinfo{volume}{368}},
  \bibinfo{pages}{489--493} (\bibinfo{year}{2020}).

\bibitem{bomfim2020predicting}
\bibinfo{author}{Bomfim, R.} \emph{et~al.}
\newblock \bibinfo{title}{Predicting dengue outbreaks at neighbourhood level
  using human mobility in urban areas}.
\newblock \emph{\bibinfo{journal}{Journal of the Royal Society Interface}}
  \textbf{\bibinfo{volume}{17}}, \bibinfo{pages}{20200691}
  (\bibinfo{year}{2020}).

\bibitem{m1925applications}
\bibinfo{author}{M'Kendrick, A.~G.}
\newblock \bibinfo{title}{{A}pplications of {M}athematics to {M}edical
  {P}roblems}.
\newblock \emph{\bibinfo{journal}{Proceedings of the Edinburgh Mathematical
  Society}} \textbf{\bibinfo{volume}{44}}, \bibinfo{pages}{98--130}
  (\bibinfo{year}{1925}).

\bibitem{kermack1991contributionsI}
\bibinfo{author}{Kermack, W.~O.} \& \bibinfo{author}{McKendrick, A.~G.}
\newblock \bibinfo{title}{Contributions to the mathematical theory of
  epidemics--{I}. 1927.}
\newblock \emph{\bibinfo{journal}{Bulletin of Mathematical Biology}}
  \textbf{\bibinfo{volume}{53}}, \bibinfo{pages}{33--55}
  (\bibinfo{year}{1991}).

\bibitem{kermack1991contributionsII}
\bibinfo{author}{Kermack, W.~O.} \& \bibinfo{author}{McKendrick, A.~G.}
\newblock \bibinfo{title}{Contributions to the mathematical theory of
  epidemics--{II}. {T}he problem of endemicity}.
\newblock \emph{\bibinfo{journal}{Bulletin of Mathematical Biology}}
  \textbf{\bibinfo{volume}{53}}, \bibinfo{pages}{57--87}
  (\bibinfo{year}{1991}).

\bibitem{kermack1991contributionsIII}
\bibinfo{author}{Kermack, W.~O.} \& \bibinfo{author}{McKendrick, A.~G.}
\newblock \bibinfo{title}{Contributions to the mathematical theory of
  epidemics--{III}. {F}urther studies of the problem of endemicity}.
\newblock \emph{\bibinfo{journal}{Bulletin of Mathematical Biology}}
  \textbf{\bibinfo{volume}{53}}, \bibinfo{pages}{89--118}
  (\bibinfo{year}{1991}).

\bibitem{diekmann2021discrete}
\bibinfo{author}{Diekmann, O.}, \bibinfo{author}{Othmer, H.~G.},
  \bibinfo{author}{Planqu{\'e}, R.} \& \bibinfo{author}{Bootsma, M. C.~J.}
\newblock \bibinfo{title}{The discrete-time {K}ermack--{McK}endrick model: {A}
  versatile and computationally attractive framework for modeling epidemics}.
\newblock \emph{\bibinfo{journal}{Proceedings of the National Academy of
  Sciences of the United States of America}} \textbf{\bibinfo{volume}{118}},
  \bibinfo{pages}{e2106332118} (\bibinfo{year}{2021}).

\bibitem{Xia_SIAM}
\bibinfo{author}{Xia, M.}, \bibinfo{author}{Greenman, C.} \&
  \bibinfo{author}{Chou, T.}
\newblock \bibinfo{title}{{PDE} models of adder mechanisms in cellular
  proliferation}.
\newblock \emph{\bibinfo{journal}{SIAM Journal on Applied Mathematics}}
  \textbf{\bibinfo{volume}{80}}, \bibinfo{pages}{1307--1335}
  (\bibinfo{year}{2020}).

\bibitem{ChildPolicy}
\bibinfo{author}{Wang, Y.}, \bibinfo{author}{Dessalles, R.} \&
  \bibinfo{author}{Chou, T.}
\newblock \bibinfo{title}{Modeling the impact of birth control policies on
  {China}'s population and age: {E}ffects of delayed births and minimum birth
  age constraints}.
\newblock \emph{\bibinfo{journal}{Royal Society Open Science}}
  \textbf{\bibinfo{volume}{9}}, \bibinfo{pages}{211619} (\bibinfo{year}{2022}).

\bibitem{SCH1984}
\bibinfo{author}{Schenzle, D.}
\newblock \bibinfo{title}{An age-structured model of pre- and post-vaccination
  measles transmission}.
\newblock \emph{\bibinfo{journal}{Mathematical Medicine and Biology}}
  \textbf{\bibinfo{volume}{1}}, \bibinfo{pages}{169--191}
  (\bibinfo{year}{1984}).

\bibitem{CAS1984}
\bibinfo{author}{Castillo-Chavez, C.} \& \bibinfo{author}{Feng, Z.}
\newblock \bibinfo{title}{Global stability of an age-structure model for {TB}
  and its applications to optimal vaccination strategies}.
\newblock \emph{\bibinfo{journal}{Mathematical Biosciences}}
  \textbf{\bibinfo{volume}{151}}, \bibinfo{pages}{135--154}
  (\bibinfo{year}{1998}).

\bibitem{RON2007}
\bibinfo{author}{Rong, L.}, \bibinfo{author}{Feng, Z.} \&
  \bibinfo{author}{Perelson, A.~S.}
\newblock \bibinfo{title}{Mathematical analysis of age-structured {HIV-1}
  dynamics with combination antiretroviral therapy}.
\newblock \emph{\bibinfo{journal}{SIAM Journal of Applied Mathematics}}
  \textbf{\bibinfo{volume}{63}}, \bibinfo{pages}{731--756}
  (\bibinfo{year}{2007}).

\bibitem{bottcher2020case}
\bibinfo{author}{B{\"o}ttcher, L.}, \bibinfo{author}{Xia, M.} \&
  \bibinfo{author}{Chou, T.}
\newblock \bibinfo{title}{Why case fatality ratios can be misleading:
  individual-and population-based mortality estimates and factors influencing
  them}.
\newblock \emph{\bibinfo{journal}{Physical Biology}}
  \textbf{\bibinfo{volume}{17}}, \bibinfo{pages}{065003}
  (\bibinfo{year}{2020}).

\bibitem{CHU2018}
\bibinfo{author}{Chuang, Y.~L.}, \bibinfo{author}{Chou, T.} \&
  \bibinfo{author}{D'Orsogna, M.~R.}
\newblock \bibinfo{title}{Age-structured social interactions enhance
  radicalization}.
\newblock \emph{\bibinfo{journal}{Journal of Mathematical Sociology}}
  \textbf{\bibinfo{volume}{42}}, \bibinfo{pages}{128--151}
  (\bibinfo{year}{2018}).

\bibitem{YAN2016}
\bibinfo{author}{Yang, J.}, \bibinfo{author}{Li, X.} \& \bibinfo{author}{Zhang,
  F.}
\newblock \bibinfo{title}{Global dynamics of a heroin epidemic model with age
  structure and nonlinear incidence}.
\newblock \emph{\bibinfo{journal}{International Journal of Biomathematics}}
  \textbf{\bibinfo{volume}{9}}, \bibinfo{pages}{1650033}
  (\bibinfo{year}{2016}).

\bibitem{LIU2019}
\bibinfo{author}{Liu, L.} \& \bibinfo{author}{Liu, X.}
\newblock \bibinfo{title}{Mathematical analysis for an age-structured heroin
  epidemic model}.
\newblock \emph{\bibinfo{journal}{Acta Applicandae Mathematicae}}
  \textbf{\bibinfo{volume}{164}}, \bibinfo{pages}{193--217}
  (\bibinfo{year}{2019}).

\bibitem{CHE2020}
\bibinfo{author}{Chekroun, A.}, \bibinfo{author}{Frioui, M.~N.},
  \bibinfo{author}{Kuniya, T.} \& \bibinfo{author}{Touaoula, T.~M.}
\newblock \bibinfo{title}{Mathematical analysis of an age structured
  heroin-cocaine epidemic model}.
\newblock \emph{\bibinfo{journal}{Discrete and Continuous Dynamical Systems -
  B}} \textbf{\bibinfo{volume}{25}}, \bibinfo{pages}{4449--4477}
  (\bibinfo{year}{2020}).

\bibitem{DIN2020}
\bibinfo{author}{Din, A.} \& \bibinfo{author}{Li, Y.}
\newblock \bibinfo{title}{Controlling heroin addiction via age-structured
  modeling}.
\newblock \emph{\bibinfo{journal}{Advances in Difference Equations}}
  \textbf{\bibinfo{volume}{521}} (\bibinfo{year}{2020}).
\newblock \urlprefix\url{https://doi.org/10.1186/s13662-020-02983-5}.

\bibitem{DUA2021}
\bibinfo{author}{Duan, X.}, \bibinfo{author}{Cheng, H.},
  \bibinfo{author}{Martcheva, M.} \& \bibinfo{author}{Yuan, S.}
\newblock \bibinfo{title}{Dynamics of an age structured heroin transmission
  model with imperfect vaccination}.
\newblock \emph{\bibinfo{journal}{International Journal of Bifurcation and
  Chaos}} \textbf{\bibinfo{volume}{31}}, \bibinfo{pages}{2150157}
  (\bibinfo{year}{2021}).

\bibitem{KHA2021}
\bibinfo{author}{Khan, A.}, \bibinfo{author}{Zaman, G.},
  \bibinfo{author}{Ullah, R.} \& \bibinfo{author}{Naveed, N.}
\newblock \bibinfo{title}{Optimal control strategies for a heroin epidemic
  model with age-dependent susceptibility and recovery-age}.
\newblock \emph{\bibinfo{journal}{AIMS Mathematics}}
  \textbf{\bibinfo{volume}{6}}, \bibinfo{pages}{1377--1394}
  (\bibinfo{year}{2021}).

\bibitem{evensen1994sequential}
\bibinfo{author}{Evensen, G.}
\newblock \bibinfo{title}{Sequential data assimilation with a nonlinear
  quasi-geostrophic model using {M}onte {C}arlo methods to forecast error
  statistics}.
\newblock \emph{\bibinfo{journal}{Journal of Geophysical Research: Oceans}}
  \textbf{\bibinfo{volume}{99}}, \bibinfo{pages}{10143--10162}
  (\bibinfo{year}{1994}).

\bibitem{gompertz1825}
\bibinfo{author}{Gompertz, B.}
\newblock \bibinfo{title}{On the nature of the function expressive of the law
  of human mortality, and on a new mode of determining the value of life
  contingencies}.
\newblock \emph{\bibinfo{journal}{Philosophical Transactions of the Royal
  Society of London}} \textbf{\bibinfo{volume}{115}}, \bibinfo{pages}{513--583}
  (\bibinfo{year}{1825}).

\bibitem{makeham1860law}
\bibinfo{author}{Makeham, W.~M.}
\newblock \bibinfo{title}{On the law of mortality and the construction of
  annuity tables}.
\newblock \emph{\bibinfo{journal}{Journal of the Institute of Actuaries}}
  \textbf{\bibinfo{volume}{8}}, \bibinfo{pages}{301--310}
  (\bibinfo{year}{1860}).

\bibitem{siler1979competing}
\bibinfo{author}{Siler, W.}
\newblock \bibinfo{title}{A competing-risk model for animal mortality}.
\newblock \emph{\bibinfo{journal}{Ecology}} \textbf{\bibinfo{volume}{60}},
  \bibinfo{pages}{750--757} (\bibinfo{year}{1979}).

\bibitem{siler1983parameters}
\bibinfo{author}{Siler, W.}
\newblock \bibinfo{title}{Parameters of mortality in human populations with
  widely varying life spans}.
\newblock \emph{\bibinfo{journal}{Statistics in Medicine}}
  \textbf{\bibinfo{volume}{2}}, \bibinfo{pages}{373--380}
  (\bibinfo{year}{1983}).

\bibitem{cohen2018gompertz}
\bibinfo{author}{Cohen, J.~E.}, \bibinfo{author}{Bohk-Ewald, C.} \&
  \bibinfo{author}{Rau, R.}
\newblock \bibinfo{title}{{G}ompertz, {M}akeham, and {S}iler models explain
  {T}aylor's law in human mortality data}.
\newblock \emph{\bibinfo{journal}{Demographic Research}}
  \textbf{\bibinfo{volume}{38}}, \bibinfo{pages}{773--841}
  (\bibinfo{year}{2018}).

\bibitem{SAMHSI2010}
\bibinfo{title}{{Substance Abuse and Mental Health Services Administration,
  Results from the 2010 National Survey on Drug Use and Health: Summary of
  National Findings. NSDUH Series H-41, HHS Publication No. (SMA) 11-4658,
  Rockville, MD}} (\bibinfo{year}{2011}).

\bibitem{Mathers2013}
\bibinfo{author}{Mathers, B.~M.} \emph{et~al.}
\newblock \bibinfo{title}{Mortality among people who inject drugs: a systematic
  review and meta-analysis}.
\newblock \emph{\bibinfo{journal}{Bulletin of the World Health Organization}}
  \textbf{\bibinfo{volume}{91}}, \bibinfo{pages}{102--123}
  (\bibinfo{year}{2013}).

\bibitem{Lindblad2016}
\bibinfo{author}{Lindblad, R.} \emph{et~al.}
\newblock \bibinfo{title}{Mortality rates among substance use disorder
  participants in clinical trials: {P}ooled analysis of twenty-two clinical
  trials within the national drug abuse treatment clinical trials network}.
\newblock \emph{\bibinfo{journal}{Journal of Substance Abuse Treatment}}
  \textbf{\bibinfo{volume}{70}}, \bibinfo{pages}{73--80}
  (\bibinfo{year}{2016}).

\bibitem{Hser2017}
\bibinfo{author}{Hser, Y.~I.} \emph{et~al.}
\newblock \bibinfo{title}{High mortality among patients with opioid use
  disorder in a large healthcare system}.
\newblock \emph{\bibinfo{journal}{Journal of Addiction Medicine}}
  \textbf{\bibinfo{volume}{11}}, \bibinfo{pages}{315--319}
  (\bibinfo{year}{2017}).

\bibitem{Sordo2017}
\bibinfo{author}{Sordo, L.} \emph{et~al.}
\newblock \bibinfo{title}{Mortality risk during and after opioid substitution
  treatment: {S}ystematic review and meta-analysis of cohort studies}.
\newblock \emph{\bibinfo{journal}{BMJ}} \textbf{\bibinfo{volume}{357}},
  \bibinfo{pages}{357:j1550} (\bibinfo{year}{2017}).

\bibitem{King2022}
\bibinfo{author}{King, C.}, \bibinfo{author}{Cook, R.},
  \bibinfo{author}{Korthuis, P.~T.}, \bibinfo{author}{Morris, C.} \&
  \bibinfo{author}{Englander, H.}
\newblock \bibinfo{title}{Causes of death in the 12 months after hospital
  discharge among patients with opioid use disorder}.
\newblock \emph{\bibinfo{journal}{Journal of Addiction Medicine}}
  \textbf{\bibinfo{volume}{16}}, \bibinfo{pages}{466--469}
  (\bibinfo{year}{2022}).

\bibitem{Bahji2020}
\bibinfo{author}{Bahji, A.}, \bibinfo{author}{Cheng, B.},
  \bibinfo{author}{Gray, S.} \& \bibinfo{author}{Stuart, H.}
\newblock \bibinfo{title}{Mortality among people with opioid use disorder: {A}
  systematic review and meta-analysis}.
\newblock \emph{\bibinfo{journal}{Journal of Addiction Medicine}}
  \textbf{\bibinfo{volume}{14}}, \bibinfo{pages}{e118--e132}
  (\bibinfo{year}{2020}).

\bibitem{gini1912variabilita}
\bibinfo{author}{Gini, C.}
\newblock \emph{\bibinfo{title}{Variabilit{\`a} e mutabilit{\`a}: {C}ontributo
  allo studio delle distribuzioni e delle relazioni statistiche}}
  (\bibinfo{publisher}{C. Cuppini}, \bibinfo{address}{Bologna, Italy},
  \bibinfo{year}{1912}).

\bibitem{xu2020diversity}
\bibinfo{author}{Xu, S.}, \bibinfo{author}{B{\"o}ttcher, L.} \&
  \bibinfo{author}{Chou, T.}
\newblock \bibinfo{title}{Diversity in biology: definitions, quantification and
  models}.
\newblock \emph{\bibinfo{journal}{Physical Biology}}
  \textbf{\bibinfo{volume}{17}}, \bibinfo{pages}{031001}
  (\bibinfo{year}{2020}).

\bibitem{Courtwright2001}
\bibinfo{author}{Courtwright, D.~T.}
\newblock \emph{\bibinfo{title}{Dark Paradise: {A} History of Opiate Addiction
  in America}} (\bibinfo{publisher}{Harvard University Press},
  \bibinfo{address}{Cambridge, MA}, \bibinfo{year}{2001}).

\bibitem{Musto1999}
\bibinfo{author}{Musto, D.~F.}
\newblock \emph{\bibinfo{title}{{The American disease: Origins of narcotic
  control}}} (\bibinfo{publisher}{Oxford University Press},
  \bibinfo{address}{Oxford, UK}, \bibinfo{year}{1999}),
  \bibinfo{edition}{3$^{rd}$} edn.

\bibitem{Courtwright1992}
\bibinfo{author}{Courtwright, D.~T.}
\newblock \bibinfo{title}{{A century of American narcotic policy}}.
\newblock In \bibinfo{editor}{Gerstein, D.~R.} \& \bibinfo{editor}{Harwood,
  H.~J.} (eds.) \emph{\bibinfo{booktitle}{Treating drug problems}},
  vol.~\bibinfo{volume}{2} (\bibinfo{publisher}{National Academies Press},
  \bibinfo{address}{Washington, DC}, \bibinfo{year}{1992}).

\bibitem{jones2019data}
\bibinfo{author}{Jones, C.~M.}, \bibinfo{author}{Warner, M.},
  \bibinfo{author}{Hedegaard, H.} \& \bibinfo{author}{Compton, W.}
\newblock \bibinfo{title}{Data quality considerations when using county-level
  opioid overdose death rates to inform policy and practice}.
\newblock \emph{\bibinfo{journal}{Drug and Alcohol Dependence}}
  \textbf{\bibinfo{volume}{204}}, \bibinfo{pages}{107549}
  (\bibinfo{year}{2019}).

\bibitem{chou2016hierarchical}
\bibinfo{author}{Chou, T.} \& \bibinfo{author}{Greenman, C.~D.}
\newblock \bibinfo{title}{A hierarchical kinetic theory of birth, death and
  fission in age-structured interacting populations}.
\newblock \emph{\bibinfo{journal}{Journal of Statistical Physics}}
  \textbf{\bibinfo{volume}{164}}, \bibinfo{pages}{49--76}
  (\bibinfo{year}{2016}).

\bibitem{anderson2001ensemble}
\bibinfo{author}{Anderson, J.~L.}
\newblock \bibinfo{title}{An ensemble adjustment {K}alman filter for data
  assimilation}.
\newblock \emph{\bibinfo{journal}{Monthly Weather Review}}
  \textbf{\bibinfo{volume}{129}}, \bibinfo{pages}{2884--2903}
  (\bibinfo{year}{2001}).

\bibitem{evensen2000ensemble}
\bibinfo{author}{Evensen, G.} \& \bibinfo{author}{Van~Leeuwen, P.~J.}
\newblock \bibinfo{title}{An ensemble {K}alman smoother for nonlinear
  dynamics}.
\newblock \emph{\bibinfo{journal}{Monthly Weather Review}}
  \textbf{\bibinfo{volume}{128}}, \bibinfo{pages}{1852--1867}
  (\bibinfo{year}{2000}).

\bibitem{nchs2021}
\bibinfo{author}{{National Center for Health Statistics}}.
\newblock \bibinfo{title}{{NCHS Fact Sheet -- June 2021: NCHS Data on Drug
  Overdose Deaths}}.
\newblock \bibinfo{howpublished}{Report} (\bibinfo{year}{2021}).

\bibitem{abuse2019mental}
\bibinfo{author}{Lipari, R.~N.} \& \bibinfo{author}{Park-Lee, E.}
\newblock \bibinfo{title}{{M}ental {H}ealth {S}ervices {A}dministration. {K}ey
  substance use and mental health indicators in the {U}nited {S}tates:
  {R}esults from the 2018 {N}ational {S}urvey on {D}rug {U}se and {H}ealth
  {(HHS Publication No. PEP19-5068, NSDUH Series H-54). Rockville, MD: Center
  for Behavioral Health Statistics and Quality}}.
\newblock \emph{\bibinfo{journal}{Substance Abuse and Mental Health Services
  Administration}}  (\bibinfo{year}{2019}).

\bibitem{brown1997introduction}
\bibinfo{author}{Brown, R.~G.} \& \bibinfo{author}{Hwang, P. Y.~C.}
\newblock \emph{\bibinfo{title}{Introduction to random signals and applied
  {K}alman filtering: with {MATLAB} exercises and solutions}}
  (\bibinfo{publisher}{Wiley}, \bibinfo{address}{Hoboken, NJ},
  \bibinfo{year}{1997}).

\bibitem{labbe2014}
\bibinfo{author}{Labbe, R.}
\newblock \bibinfo{title}{{K}alman and {B}ayesian {F}ilters in {P}ython,
  \url{https://github.com/rlabbe/Kalman-and-Bayesian-Filters-in-Python/blob/master/Appendix-E-Ensemble-Kalman-Filters.ipynb}}
  (\bibinfo{year}{2022}).

\bibitem{brown2012introduction}
\bibinfo{author}{Brown, R.~G.} \& \bibinfo{author}{Hwang, P. Y.~C.}
\newblock \emph{\bibinfo{title}{Introduction to random signals and applied
  {K}alman filtering: with {MATLAB} exercises and solutions}}
  (\bibinfo{publisher}{Wiley}, \bibinfo{address}{Hoboken, NJ},
  \bibinfo{year}{2012}).

\end{thebibliography}
\newpage
\section*{Methods}
\appendix
\subsection*{Age-structured overdose model}
\begin{figure}
    \centering
    \includegraphics{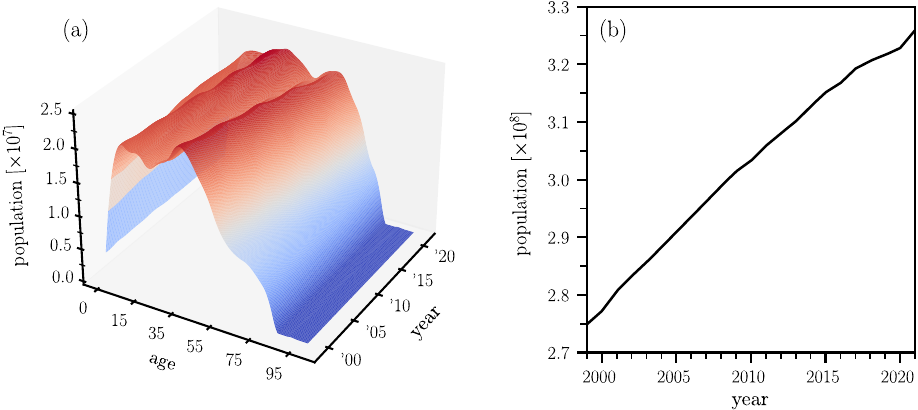}
    \caption{\textbf{Age-structured United States population data.}
      (a) Interpolated age-structured population data. The
      interpolation is based on bivariate splines of degree 2. (b) The
      overall United States population increases almost linearly
      between 1999 and 2021.}
    \label{fig:N(a,t)}
\end{figure}
The mathematical model we use to describe the age-stratified evolution of SUD cases is given by
\begin{equation}
 \left[\frac {\partial} {\partial a}  + \frac {\partial} {\partial t} \right] 
 n(a,t)  = - \mu(a, t)  n(a, t) + r(a,t) [N(a,t) - n(a,t) ]\,,
\label{eq:pde_model}
\end{equation}
where $n(a,t) \mathrm{d}a$ is the population affected by SUD
with age between $a$ and $a + \mathrm{d}a$ at time $t$. The associated mortality 
is $\mu(a,t)$ and $r(a,t)$ is the influx rate of new SUD cases from
$N(a,t)-n(a,t)$, the pool of individuals without SUD. Finally, $N(a,t)$ is the general 
population with age between $a$ and $a +
\mathrm{d}a$ at time $t$.  We set the initial age and time
$a_0=t_0=0$. The initial distribution of SUD cases is given by $n(a,
t=0) = \rho(a)$. We also set $n(a=0, t) =0$ such that no population of
age $a=0$ exists at any time.  We solve Eq.~\eqref{eq:pde_model} using
the method of characteristics and distinguish the two cases $a \geq t$
and $a < t$.  For $a \geq t$, the characteristic begins at $t=0$ and
$n(a,t)$ will remain constant along $a = t$, yielding
\begin{align}
\begin{split}
n(a,  t)  &= \rho (a -t) e^{-\int_{0}^t \mu(s+a-t, s)+r(s+a-t,s)\, \mathrm{d}s}\, \\
&+\int_{0}^{t} r(s + a - t, s)N(s + a - t, s)e^{-\int_{s}^t \mu(z+a-t, z)+r(z+a-t, z) \mathrm{d}z}\, \mathrm{d}s
\quad(a \geq t)\,.
\label{analysol1}
\end{split} 
\end{align}
For $a < t$, the characteristic will begin at $a
= 0$ and $n(a,t)$ will remain constant along $t = a$ so that
\begin{align}
\begin{split}
n(a,  t)  = \int_{0}^{a} r(s, s -a +t)N(s,s -a +t )e^{-\int_{s}^a \mu(z, z - a + t)+r(z, z - a + t)\, \mathrm{d}z}\, \mathrm{d}s \quad (a < t )\,.
\end{split}
\label{analysol2}
\end{align}
We write the mortality rate $\mu(a,t)$ as the sum of a baseline
mortality rate, $\mu_0(a,t)$, and a drug-caused excess mortality rate,
$\mu_{\rm d}(a,t)$, so that $\mu(a,t)=\mu_0(a,t)+{\mu}_{\rm
  d}(a,t)$. The baseline mortality $\mu_0(a,t) = \mu_0(a)$ is derived from the Gompertz--Makeham--Siler
mortality model for human death~\cite{gompertz1825,makeham1860law,siler1979competing,siler1983parameters,cohen2018gompertz}
and is assumed to be time-independent, implying that age-stratified mortality has not changed appreciably over the past twenty years. 
The quantity $\mu_{\rm d}(a,t) = \mu_{\rm d}$ is instead assumed to be age-independent and inferred from data, so that we effectively 
neglect any time dependence over a data assimilation cycle of one year. Within a data assimilation window of one year, we thus set
\begin{equation}
    \mu(a,t)= \mu_0(a) + \mu_{\rm d}= \gamma_1 e^{-\lambda_1 a}+\gamma_2+\lambda_2 e^{\lambda_2(a-M)}+\mu_{\rm d} \quad (\gamma_{1,2},\lambda_{1,2},M>0)\,,
\end{equation}
\begin{figure}
    \centering
    \includegraphics{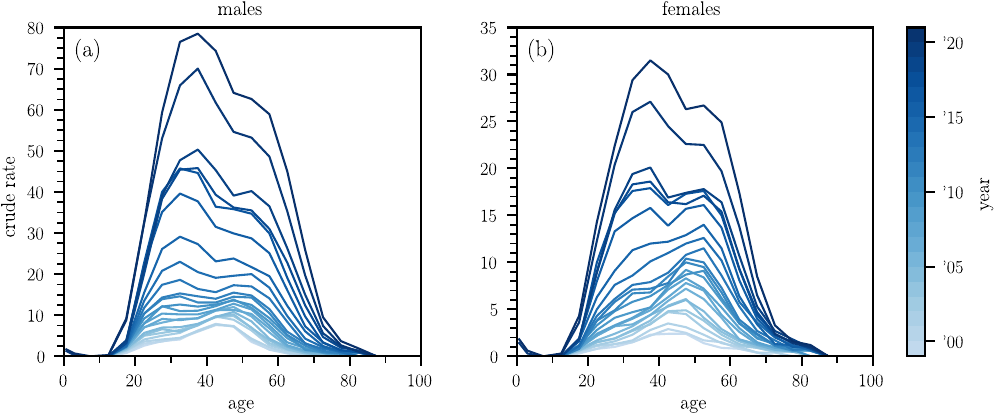}
    \caption{\textbf{Age-structured overdose crude rates in the United
        States.} Age-structured overdose crude rates in the United
      States, distinguished by gender: (a) males and (b) females.}
    \label{fig:crude_rates}
\end{figure}
where  $\gamma_1=0.00258/\mathrm{year}$, $\gamma_2=0.00037/\mathrm{year}$,
$\lambda_1=5.09657/\mathrm{year}$, $\lambda_2=0.09040/\mathrm{year}$,
and $M=83.22956~\mathrm{year}$ in accordance with Ref.\,\cite{cohen2018gompertz}. 
Since the majority of drug overdose fatalities occur among
males~\cite{nchs2021,dorsogna2023fentanyl} as seen in Fig.~\ref{fig:crude_rates}
we explicitly used parameters pertaining to males in the United States. Parameters vary slightly from year to year, 
we selected the ones for 2010.  It is worth noting that we conducted a sensitivity analysis through 
additional simulations using variations on the above choices for $\gamma_1, \gamma_2, 
\lambda_1, \lambda_2, M$ including values 
specific to females in the United States. We find that the choice of these
parameters does not substantially affect our results.

We allow the parameter ${\mu}_{\rm d}$ to change from one year to the
next. This choice is dictated by the CDC WONDER database providing
yearly lists of overdose deaths, although monthly updates could also be implemented. To describe an
age-dependent influx into the pool of SUD users, we set $r(a,t)\equiv
r(a)=\left[r_1 f (a;\alpha_1,\beta_1)+r_2
  f (a;\alpha_2,\beta_2)\right]/2$, where $f(a; \alpha, \beta)  = \beta^{\alpha}/ \Gamma(\alpha) \, a^{\alpha -1} e^{-\beta a} $
   is the gamma
distribution with shape and rate parameters $\alpha$ and
$\beta$.  The maximum of a gamma function is given at the
age $a^{\rm max} = (\alpha -1) /\beta$.  The evolution of $n(a,t)$ in data
assimilation cycles requires us to evaluate the derivative of
Eq.~\eqref{analysol1} w.r.t.\ $t$.  Given the two variable function $g(x,y)$, we use the notation $g^\prime(x,y)$ and $g_{\prime}(x,y)$ to
indicate a partial derivative in the first and second argument of $g$, respectively.
For $a \geq t$, the rate of change of $n(a,t)$ can thus be written as
\begin{align}
\begin{split}
\frac{\partial n(a,t)}{\partial t}=&-\left\{\rho' (a-t)+\rho(a-t)\left[\gamma_1 e^{-\lambda_1(a-t)}+\gamma_2+\lambda_2 e^{\lambda_2(a-t-M)}+{\mu}_{\rm d}+r(a-t)\right]\right\} \\
&\times e^{\gamma_1/\lambda_1 e^{-\lambda_1 a} (1-e^{\lambda_1 t})-e^{\lambda_2 (a-M)}\left(1-e^{-\lambda_2 t}\right)-(\gamma_2+{\mu}_{\rm d}) t-\int_{0}^t r(s+a-t)\, \mathrm{d}s}+r(a)N(a,t)\\
&-\int_0^t e^{\gamma_1/\lambda_1 e^{-\lambda_1 a} (1-e^{\lambda_1(t-s)})-e^{\lambda_2 (a-M)}\left(1-e^{-\lambda_2 (t-s)}\right)-(\gamma_2+{\mu}_{\rm d}) (t-s)-\int_{s}^t r(z+a-t)\, \mathrm{d}z} N(s+a-t,s)\times \\
& \hspace{2.5em}\left\{r(s+a-t)\left[\gamma_1 e^{-\lambda_1(s+a-t)}+\gamma_2+\lambda_2 e^{\lambda_2(s+a-t-M)}+{\mu}_{\rm d}+r(s+a-t)\right]+r'(s+a-t)\right\}\,\mathrm{d}s\\
&-\int_0^t  e^{\gamma_1/\lambda_1 e^{-\lambda_1 a} (1-e^{\lambda_1(t-s)})-e^{\lambda_2 (a-M)}\left(1-e^{-\lambda_2 (t-s)}\right)-(\gamma_2+{\mu}_{\rm d}) (t-s)-\int_{s}^t r(z+a-t)\, \mathrm{d}z} \times \\
&\hspace{2.5em}N'(s + a - t, s) r(s + a - t)\,\mathrm{d}s\,.
\end{split}
\end{align}
For $a<t$ we obtain
\begin{align}
\frac{\partial n(a,t)}{\partial t}=\int_{0}^{a} r(s) N_{\prime}(s,s-a+t) e^{\gamma_1/\lambda_1 (e^{-\lambda_1 a}-e^{-\lambda_1 s})-e^{-\lambda_2 M}(e^{\lambda_2 a}-e^{\lambda_2 s})-(\gamma_2+{\mu}_{\rm d})(a-s)-\int_{s}^a r(z)\, \mathrm{d}z}\, \mathrm{d}s\,.
\end{align}
The integrals $\int_{s}^t r(z+a-t)\, \mathrm{d}z$, $\int_{0}^t
r(s+a-t)\, \mathrm{d}s$, and $\int_{s}^a r(z)\, \mathrm{d}z$ can be
evaluated using the identity
\begin{align}
\begin{split}
\int_s^t \frac{\beta^\alpha}{\Gamma(\alpha)} (z+a-t)^{\alpha-1}e^{-\beta(z+a-t)}\,\mathrm{d}z=\frac{1}{\Gamma(\alpha)}\left[\Gamma(\alpha,(a - t +s)\beta)-\Gamma(\alpha,a\beta)\right]\,,
\end{split}
\end{align}
where $\Gamma(s,x)=\int_x^\infty t^{s-1}e^{-t}\,\mathrm{d}t$ denotes
the upper incomplete gamma function. We evaluate the remaining
integrals $\int_0^t (\cdot) \,\mathrm{d}s$ numerically.
Finally, the initial condition used to solve Eq.~\eqref{eq:pde_model} and to
obtain the simulation results in Figs.~\ref{fig:US_forecast},  \ref{fig:future_forecast}, 
and \ref{fig:county_forecast} is 
\begin{equation}
 \rho(a) = 0.015 N_0 f(a;\alpha_{0},\beta_{0}). 
\end{equation}
To obtain the curves shown in Figs.~\ref{fig:US_forecast} and \ref{fig:future_forecast} 
we set $N_0=274,886,150$, the population of the United States
between ages 0--85 in 1999. We also select $f(a;\alpha_{0},\beta_{0})$ to be a gamma distribution with shape and rate
parameters $\alpha_0$ and $\beta_0$, chosen as
$\alpha_0=12$ and $\beta_0=1/(3~\mathrm{year})$ such
that the maximum of the distribution is at $a_{\rm max} =  33$ years. The prefactor of 0.015 is
chosen such that initially 1.5\% of the population are suffering from
an SUD consistent with corresponding survey
data~\cite{abuse2019mental, SAMHSI2010}.
To obtain the curves in the county-level analysis, we initially set $N_0$ to match the respective population sizes between ages 0--85 for the years 1999 (Los Angeles County) and 2013 (Cook County and New York City). Specifically, $N_0$ was set to $9,437,290$ for Los Angeles County, $8,405,837$ for New York City, and $5,240,700$ for Cook County. 
We also set $\alpha=17$ (Los Angeles County) and $\alpha=12$ (Cook County and New York City) and 
used the same value of $\beta$ as in the national
analysis. 
\subsection*{Interpolating population data}
We infer the age-structured population function $N(a,t)$ from
nationwide population data that is available from the CDC WONDER
database. In Fig.~\ref{fig:N(a,t)}(a) we show an interpolated and
differentiable population function $N(a,t)$. In all simulations, we use interpolations that are based on bivariate splines of degree 2. In Fig.~\ref{fig:N(a,t)}(b), we show the almost linear increase of the population of the United States between ages 0-85 from 1999 to 2021.
\subsection*{Ensemble Kalman filter}
\begin{figure}
    \centering
    \includegraphics[width=0.8\textwidth]{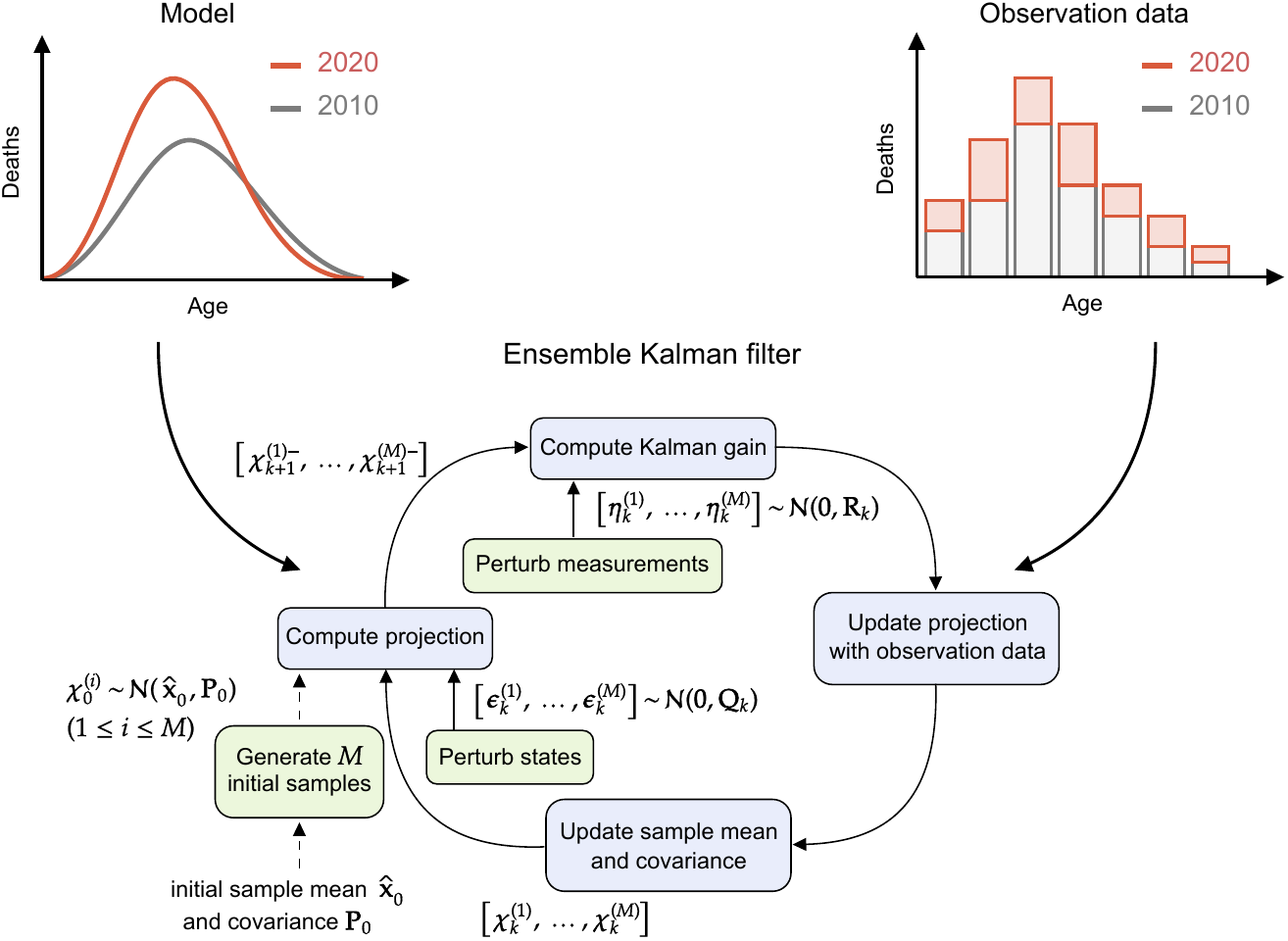}
    \caption{\textbf{Ensemble Kalman filter schematic.} 
      We use an EnKF to combine a mechanistic model of drug-overdose
      fatalities with corresponding observational data. Blue boxes show
      the main steps (\ie, projection and update) in an EnKF
      cycle. Green boxes represent the initial sample generation
      process and perturbations that are added during the projection
      and update steps. The schematic is adapted from
      \cite{brown1997introduction}.}
    \label{fig:enkf_schematic}
\end{figure}
To combine the age-structured drug overdose model \eqref{eq:pde_model}
with corresponding observational data, we use an ensemble Kalman filter
(EnKF)~\cite{evensen1994sequential} as implemented in
\cite{labbe2014,bottcher2023forecasting}. Figure~\ref{fig:enkf_schematic}
shows a schematic of the main EnKF steps.

In accordance with \cite{brown2012introduction,crassidis2004optimal},
the evolution of any system state $\mathbf{x}(t)$ (\eg, number of SUD
cases, mortality and addiction rates) and observed state
$\mathbf{z}(t)$ (\eg, number of overdose fatalities) are described by
the stochastic differential equations
\begin{align}
\begin{split}
\dot{\mathbf{x}}&=\mathbf{f}(\mathbf{x},t)+\mathbf{w}(t)\quad\mathbf{w}(t)\sim\mathcal{N}(\mathbf{0},\mathbf{Q}(t))\\
\mathbf{z}&=\mathbf{h}(\mathbf{x},t)+\mathbf{v}(t)\quad\mathbf{v}(t)\sim\mathcal{N}(\mathbf{0},\mathbf{R}(t))
\end{split}\,,
\label{eq:state_space_model}
\end{align}
where $\mathbf{Q}(t)$ and $\mathbf{R}(t)$ denote the covariance
matrices associated with the Gaussian process noise
$\mathcal{N}(\mathbf{0},\mathbf{Q}(t))$ and Gaussian measurement noise
$\mathcal{N}(\mathbf{0},\mathbf{R}(t))$ at time $t$, respectively. We
assume the quantities $\mathbf{Q}(t)$ and $\mathbf{R}(t)$ to be
given. The function $\mathbf{f}(\cdot)$ describes the dynamics of the
system state $\mathbf{x}(t)$, while $\mathbf{h}(\cdot)$ maps
$\mathbf{x}(t)$ to a measurable quantity. Both functions can be
non-linear.

For the specific case of our age-structured model defined in
Eq.\,\eqref{eq:pde_model}, element $x_j(t)$ of the state vector
$\mathbf{x}(t)$ corresponds to $n(a_j,t)\equiv n(a_0+(j-1)\Delta a,t)$
($j\in\{1,\dots,N_a\}$), the density of individuals whose age lies within
the $[a_0+(j-1)\Delta a, a_0+j\Delta a)$ interval at time $t$. Here,
  $N_a$ and $\Delta a$ denote the number of discretizations of the age
  interval and the corresponding age discretization step,
  respectively. Thus, we write
\begin{equation}
\mathbf{x}(t)=[n(a_1,t),n(a_2,t),\dots]^\top\,.
\label{eq:unaugmented}
\end{equation}
For the numerical solution of Eq.~\eqref{eq:state_space_model}, we
also discretize the simulation time interval $[0,T]$ into $N_t$
equidistant intervals of duration $\Delta t=T/N_t$. In all of our simulations, we fixed $\Delta t$ at 0.1. To combine the mechanistic model in Eq.\,\eqref{eq:pde_model} with
empirical data on overdose fatalities, we augment the system state
\eqref{eq:unaugmented} by
\begin{equation}
\tilde{D}(a_j,t)=\int_0^t\mu(a_j,\tilde{t})n(a_j,\tilde{t})\,\mathrm{d}\tilde{t}\quad(j\in\{1,\dots,N_a\})\,,
\end{equation}
where $\tilde{D}(a_j,t)$ is the cumulative number of overdose deaths
in the age interval $[a_{j-1},a_{j})$ up to time $t$. To avoid dealing with large differences between predicted and observed fatalities in our numerical calculations, we normalize both quantities by dividing them by 1,000, thereby measuring overdose deaths per 1,000 individuals in the system state. Since we wish to
  estimate model parameters such as $\mu_{\rm d}$ and
  $r_1,r_2,\alpha_1,\beta_1,\alpha_2,\beta_2$, we also augment the
  system state \eqref{eq:unaugmented} by the log-transforms
  $\tilde{\mu}_{\rm d}=\log(\mu_{\rm d})$, $\tilde{r}_1=\log(r_1)$,
  $\tilde{r}_2=\log(r_2)$, $\tilde{\alpha}_1=\log(\alpha_1)$,
  $\tilde{\beta}_1=\log(\beta)_1$, $\tilde{\alpha}_2=\log(\alpha_2)$
  and $\tilde{\beta}_2=\log(\beta_2)$. Therefore, the final augmented
  system state is
\begin{align}
\mathbf{x}(t)=\Big[&n(a_1,t),\dots,n(a_{N_a},t),\tilde{D}(a_1,t),\dots,\tilde{D}(a_{N_a},t), \tilde{\mu}_{\rm d},\tilde{r}_1,\tilde{r}_2,\tilde{\alpha}_1,\tilde{\beta}_1,\tilde{\alpha}_2,\tilde{\beta}_2\Big]^{\top}\,.
\label{eq:augmented}
\end{align}
Prior to each prediction step, we apply an exponential transform to
render the parameters $\mu_{\rm d}$ and
$r_1,r_2,\alpha_1,\beta_1,\alpha_2,\beta_2$ positive and avoid sign
changes.
To accurately solve the evolution of $n(a,t)$ numerically, we must use
a sufficiently large number of age windows $N_a$ in our
simulations. However, since the age windows in our simulations are
more granular than those available from overdose fatality data, we
apply a coarse-graining procedure. The nationwide CDC WONDER data we
utilized has 22 age groups with $a'_0=0,a'_{22}=120$ and $\Delta
a'_1=1,\Delta a'_2=4,\Delta a'_3=5,\dots,\Delta a'_{21} = 5,\Delta
a'_{22}=20$ years. To differentiate between the age discretization in
the observational data and the age discretization in the underlying
model, we employ a superscript $'$ notation. In the county-level
forecasts, we employed 10-year age groups, effectively reducing the
number of age groups from 22 to 11. The following discussion of the
EnKF parameterization and implementation will be based on 22 age
groups, but the same considerations also apply to the county-level
analysis with fewer age groups.

To reduce granularity and combine the modeled quantities
$\tilde{D}(a_j,t)$ with corresponding observational data, we numerically
integrate $\tilde{D}(a_j,t)$ over the age windows
$[a'_{\ell-1},a'_{\ell})$ ($\ell\in\{1,\dots, 22\}$) to obtain the
  cumulative number of deaths $D(a'_\ell,t)$ in this age interval at
  time $t$. Here, $a'_\ell=a'_0+\sum_{m=1}^{\ell}\Delta a_{m}$ for
  $\ell\geq 1$. Based on the described mapping of $\tilde{D}(a_j,t)$
  to $D(a'_\ell,t)$, the measurement function becomes
\begin{equation}
\mathbf{h}(\mathbf{x}(t))=[D(a'_1,t),D(a'_2,t),\dots]^\top\,.
\end{equation}
In our simulations, we set the initial values
$n(a_j,0)=\tilde{D}(a_j,0)=0$. In the nationwide analysis, we initially set $\mu_{\rm d}=2\times 10^{-3}\mathrm{/year}$, $r_1=r_2=2\times
10^{-2}\mathrm{/year}$, $\alpha_1=10$, $\beta_1=1/(3~\mathrm{year})$,
$\alpha_2=15$, and $\beta_2=1/(3~\mathrm{year})$. We have chosen the initial values
of $r_1,r_2$ in accordance with corresponding empirical data on the
number of substance initiates~\cite{abuse2019mental, SAMHSI2010}
The initial maximum values of the gamma distributions
$f (a; \alpha_1, \beta_1), f(a; \alpha_2, \beta_2) $ are attained at ages  
$a^{\rm max}_1 = (\alpha_1 -1)/\beta_1=27$  years and $a^{\rm max}_2 = (\alpha_2 -1) /\beta_2=42$ years, respectively.
All initial covariances are set to $10^{-4}$, except for the diagonal elements
associated with the log-transforms of $\mu_{\rm d}$, $r_1,r_2$ and
$\alpha_1$, $\beta_1$, $\alpha_2$, $\beta_2$, which are set to $1$,
respectively. Process and observation noise covariances are assumed to
be time-independent and given by $\mathbf{Q}=10^{-4}J_{2N_a+7}$ and
$\mathbf{R}=\mathrm{diag}(2\times 10^{-3},\dots,2\times 10^{-3})$,
respectively. Here, $J_n$ denotes the $n\times n$ matrix of ones. Our
simulations run from the beginning of 1999 until the end of 2024. The age discretization is
$\Delta a = 1.2$ years with $a_0=0$ and $a_{N_a}=120$ years. Population data is available for ages between 0 and 85. However, since
overdose fatalities in groups below 10 years and above 70 years are
statistically insignificant, we truncated the system state
accordingly. To align model parameters with initial observational data,
we performed two full data assimilation cycles for the first year
(1999) before starting the main forecasting algorithm that produces
forecasts for all years from 1999 to 2024. The number of EnKF
ensemble members is $M=10^3$.

We used different initial values for the county-level forecasts.
For Los Angeles County, the initial values were set as follows: $\mu_{\rm d}=2.5\times 10^{-3}\mathrm{/year}$, $r_1=r_2=2\times
10^{-2}\mathrm{/year}$, and $\alpha_1=\alpha_2=17$. For Cook County, the values were set as $\mu_{\rm d}=5\times 10^{-3}\mathrm{/year}$, $r_1=r_2=6\times
10^{-2}\mathrm{/year}$, $\alpha_1=8$, and $\alpha_2=15$. Lastly, for New York City, the values were set as $\mu_{\rm d}=5\times 10^{-3}\mathrm{/year}$, $r_1=r_2=6\times
10^{-2}\mathrm{/year}$, $\alpha_1=8$, and $\alpha_2=17$. The initial values of $\beta_1$ and $\beta_2$ are as in the nationwide analysis. To account for the smaller number of overdose fatalities
at the county level, we adjust the process and observational noise
matrix elements to have values of the order $10^{-8}-10^{-7}$. The
number of EnKF ensemble members is $M=10^3$ (Los Angeles County) and
$M=10^2$ (Cook County and New York City).

At every time point $t$, we use the EnKF to determine the state
posterior distribution given all prior observations. Before starting
the data assimilation procedure, we generate an initial ensemble
$[\boldsymbol{\chi}_0^{(1)},\dots,\boldsymbol{\chi}_0^{(M)}]$ that
consists of $M$ ensemble members
$\boldsymbol{\chi}_0^{(i)}\sim\mathcal{N}(\hat{\mathbf{x}}_0,\mathbf{P}_0)$
($i\in\{1,\dots, M\}$). The quantities $\hat{\mathbf{x}}_0$ and
$\mathbf{P}_0$ denote the given initial state and covariance
estimates, respectively.

To perform forecast and update iterations using a Kalman filter, one
uses state estimates $\boldsymbol{\chi}_{k}^{(i)}$ at time $t_k$ to
calculate predicted state estimates $\boldsymbol{\chi}_{k+1}^{(i)-}$
at time $t_{k+1}$. These predicted state estimates are then combined
with observational data to obtain an updated state
$\boldsymbol{\chi}_{k+1}^{(i)}$. We use the superscript ``$-$'' in
$\boldsymbol{\chi}_{k+1}^{(i)-}$ to distinguish between predicted
(\ie, prior) state estimates and updated (\ie, posterior) state
estimates. In the remainder of this section, we describe the two main EnKF steps:
(i) forecasting the evolution of the system state and (ii) updating
the predicted state estimates using observational data. We use the
shorthand notation $y_k\equiv y(t_k)$ to refer to a quantity $y$ at
time $t_k=k \Delta t$ ($k\in\{0,\dots, N_t\}$).

\begin{enumerate}
    \item[(i)] \textbf{Forecast Step:} For each ensemble member, the predicted state estimate $\boldsymbol{\chi}_{k+1}^{(i)-}$ at time $t_{k+1}$ is given by 
    \begin{equation}
    \boldsymbol{\chi}_{k+1}^{(i)-}=\boldsymbol{\chi}_{k}^{(i)}+\Delta t\,\mathbf{f}(\boldsymbol{\chi}_{k}^{(i)},t_k)+\boldsymbol{\epsilon}^{(i)}_{k}\,,    
    \end{equation}
    where $\boldsymbol{\epsilon}^{(i)}_{k}\sim\mathcal{N}(\mathbf{0},\mathbf{Q}_k)$ models Gaussian process noise. Using the predicted state estimates $\hat{\mathbf{x}}_{k+1}^-$, we compute the corresponding ensemble mean, $\hat{\mathbf{x}}_{k+1}$, and covariance matrix, $(\mathbf{P}_{\hat{\mathbf{x}} \hat{\mathbf{x}}}^-)_{k+1}$, according to
    \begin{widetext}
    \begin{align}
    \hat{\mathbf{x}}_{k+1}^-&=\frac{1}{M}\sum_{i=1}^M\boldsymbol{\chi}_{k+1}^{(i)-}\\
    (\mathbf{P}_{\hat{\mathbf{x}}\hat{\mathbf{x}}}^-)_{k+1}&=\frac{1}{M-1}\sum_{i=1}^M\left[\boldsymbol{\chi}_{k+1}^{(i)-}-\hat{\mathbf{x}}_{k+1}^-\right] \left[\boldsymbol{\chi}_{k+1}^{(i)-}-\hat{\mathbf{x}}_{k+1}^-\right]^\top\label{eq:P_xx}\,.
    \end{align}
    \end{widetext}
    Although the covariance matrix $(\mathbf{P}_{\hat{\mathbf{x}}\hat{\mathbf{x}}}^-)_{k+1}$ is not required in the EnKF iteration, it is useful to estimate confidence intervals of $\hat{\mathbf{x}}_{k+1}^-$.
    \item[(ii)] \textbf{Update Step:} We first compute the ensemble mean of the predicted observation 
    \begin{equation}
    \hat{\mathbf{z}}_{k+1}^-\equiv\frac{1}{M}\sum_{i=1}^M\mathbf{z}_{k+1}^{(i)-}=\frac{1}{M}\sum_{i=1}^M\mathbf{h}(\boldsymbol{\chi}_{k+1}^{(i)-})
    \end{equation}
    as well as the corresponding covariances
    
    \begin{widetext}
    \begin{align}
    \begin{split}
    (\mathbf{P}_{\hat{\mathbf{z}}\hat{\mathbf{z}}}^-)_{k+1}&=\frac{1}{M-1}\sum_{i=1}^M\left[\mathbf{h}(\boldsymbol{\chi}_{k+1}^{(i)-})-\hat{\mathbf{z}}_{k+1}^- \right]\left[\mathbf{h}(\boldsymbol{\chi}_{k+1}^{(i)-})-\hat{\mathbf{z}}_{k+1}^-\right]^\top + \mathbf{R}_{k+1}\\
    (\mathbf{P}_{\hat{\mathbf{x}}\hat{\mathbf{z}}}^-)_{k+1}&=\frac{1}{M-1}\sum_{i=1}^M\left[\boldsymbol{\chi}_{k+1}^{(i)-}-\hat{\mathbf{x}}_{k+1}^-\right]\left[\mathbf{h}(\boldsymbol{\chi}_{k+1}^{(i)-})-\hat{\mathbf{z}}_{k+1}^-\right]^\top\,.
    \end{split}
    \end{align}
    \end{widetext}
    
    \noindent
    The Kalman gain is
    \begin{equation}
    \mathbf{K}_{k+1}=(\mathbf{P}_{\hat{\mathbf{x}}\hat{\mathbf{z}}}^-)_{k+1}  (\mathbf{P}_{\hat{\mathbf{z}}\hat{\mathbf{z}}}^-)_{k+1}^{-1}\,.
    \end{equation}
    For a given observation $\mathbf{z}_{k+1}$, the state update of ensemble member $i$ is
    \begin{equation}
    \boldsymbol{\chi}_{k+1}^{(i)}=\boldsymbol{\chi}_{k+1}^{(i)-}+\mathbf{K}_{k+1}\left[\mathbf{z}_{k+1}+\boldsymbol{\eta}^{(i)}_{k+1}-\mathbf{h}(\boldsymbol{\chi}_{k+1}^{(i)-})\right]\,,
    \end{equation}
    where $\boldsymbol{\eta}^{(i)}_{k+1}\sim\mathcal{N}(\mathbf{0},\mathbf{R}_{k+1})$ models Gaussian measurement noise. 
    Finally, the updated state estimate and the corresponding covariance matrix are
    \begin{align}
    \begin{split}
    \hat{\mathbf{x}}_{k+1}&=\frac{1}{M}\sum_{i=1}^M \boldsymbol{\chi}_{k+1}^{(i)}\\
    (\mathbf{P}_{\hat{\mathbf{x}}\hat{\mathbf{x}}})_{k+1}&=(\mathbf{P}_{\hat{\mathbf{x}}\hat{\mathbf{x}}}^-)_{k+1}-\mathbf{K}_{k+1}(\mathbf{P}_{\hat{\mathbf{z}}\hat{\mathbf{z}}}^-)_{k+1}\mathbf{K}_{k+1}^\top\,.
    \end{split}
    \end{align}
\end{enumerate}
During each update step, we assign the entry in each ensemble member
that corresponds to the logarithm of the drug-caused mortality rate,
$\tilde{\mu}_{\rm d}=\log(\mu_{\rm d})$, to be equal to the logarithm
of the ratio of observed overdose fatalities and the ensemble mean of
the population, while also accounting for the underlying noise.
\subsection*{Overdose mortality data}
Overdose fatalities extracted from the CDC WONDER database were
identified using the International Classification of Diseases, Tenth
Revision (ICD--10) cause-of-death codes X40--44 (unintentional), X60--64
(suicide), X85 (homicide), Y10--14 (undetermined intent), and all other
drug-induced causes. National and county level data were extracted for
the 1999--2020 period. The drug categories examined are poisoning by
narcotics and psychodysleptics (hallucinogens) (T40) and by
psychostimulants with abuse potential (T43.6). Specific subcategories
analyzed within T40 are heroin (T40.1), natural and semisynthetic
opioids (T40.2), and synthetic opioids other than methadone
(T40.4). Deaths involving more than one drug type were included in
each applicable category. Entries with an insufficient number of
deaths were excluded.

\end{document}